# QUASAR-GALAXY CLUSTERING THROUGH PROJECTED GALAXY COUNTS AT Z=0.6-1.2

Shaohua Zhang[1,2], TingGui Wang[2], Huiyuan Wang[2], Hongyan Zhou[1,2]

## ABSTRACT

We investigate the spatial clustering of galaxies around quasars at redshifts from 0.6 to 1.2 using the photometric data from Sloan Digital Sky Survey (SDSS) Stripe 82 . The quasar and galaxy cross-correlation functions are measured through the projected galaxy number density $n(r_p)$ on scales $0.05 < r_p < 20\ h^{-1}$Mpc around quasars for a sample of 2300 quasars from Schneider et al. (2007). We detect strong clustering signals at all redshifts, and find that the clustering amplitude increases significantly with redshift. We examine the dependence of the quasar-galaxy clustering on quasar and galaxy properties and find that the clustering amplitude is significantly larger for quasars with more massive black holes, or with bluer colors, while the dependence on quasar luminosity is absent. We also show that quasars have a stronger correlation amplitude with blue galaxies than with red galaxies. We finally discuss the implication of our finding.

*Subject headings:* galaxies:clusters - large-scale structure of universe - quasars: general

## 1. INTRODUCTION

Evidences have been mounted that in the local universe most massive galaxies host supermassive black holes (SMBHs) at their centers (Kormendy & Richstone 1995; Gebhardt et al. 2000; Merritt & Ferrarese 2001). These black holes are the relics of past nuclear activity, i.e., growth via accretion of gas in the galaxy (e.g., Soltan 1982; Small & Blandford 1992; Yu & Tremaine 2002). The strong link between black holes and their host galaxies, as revealed by correlations between black hole mass and the stellar velocity dispersion or the mass of the host galaxy bulge ($M_{\rm BH}-\sigma$, $\rm M_{BH}-M_b$ relation) in local galaxies (e.g., Magorrian et al. 1998; Gebhardt et al. 2000; Ferrarese & Merritt 2000; Merritt & Ferrarese 2001; Ueda et al. 2003), implies the co-evolution of the black hole with its host galaxy. This is further sustained by the similar redshift evolution of overall star formation and quasars luminosity function. However, it remains unclear what drives this co-evolution. Important issues continue to be poorly understood, such as what triggers the gas fueling, and how nuclear activity affects the subsequent evolution of their host galaxies. The answers to these questions must be lied in various correlations between active black holes and their host galaxies as well as the environment of their hosts. But it is difficult to directly study the host galaxy of a quasar, even at a quite moderate redshift, because quasar outshines its host by a large factor.

Quasar clustering and quasar-galaxy cross-correlation provide a very effective way to quantify both the environment effects and their host properties in a statistical sense. The current cosmological models show that structures (dominated by dark matter) in the universe grow from small primordial density fluctuations, detected in the cosmic microwave background, in the early universe mainly via gravitational interaction. Galaxies populate in the collapsed dark matter haloes (DMHs), and their properties are expected to be closely related to the mass of DMHs (e.g., Scoccimarro et al. 2001; Berlind & Weinberg 2002; Yang, Mo & van den Bosch 2003; Kravtsov et al. 2004; Zheng et al. 2005). In other words, galaxies trace DMHs. By measuring the clustering properties of quasars and galaxies, one can deduce what dark matter halo may host a quasar. The cross-correlation of quasars and normal galaxies can tell us how the quasars and galaxies are physically related (e.g., Hopkins et al. 2007). These relations can also be used to constrain the duty cycles of nuclear activity (e.g., Shankar, Weinberg, & Shen 2010).

There are two different approaches to measure such clustering. The first one is counting the excessive surface number density of galaxies at various distances from quasars. This over-density has a simple relation to the angular cross-correlation function between quasars and galaxies. The angular correlation function is widely used in the characterization of the large scale structure of the universe. This method works even in the absence of the redshift information of galaxies. On the other hand, if redshifts of galaxies are known, there is a more precise way to describe the clustering properties using the two-point correlation function (2PCF) in three dimension space.

At low redshifts, the cross-correlation between active glactic nuclei (AGNs) and galaxies are well studied. Based on analysis of the angular cross-correlation function between AGNs and galaxies or the overdensity of galaxies around AGNs, earlier works indicated that quasars and Seyfert galaxies are located in rich cluster of galaxies (e.g., Bahcall et al. 1969; Yee & Green 1984, 1987; Ellingson et al. 1991; Laurikainen & Salo 1995; De Robertis, Yee, & Hayhoe 1998; Smith et al. 1995, 2000). However, recent studies of large low redshift AGN samples from the Sloan Digital Sky Survey (SDSS) or 2 degree Field (2dF) survey suggested that AGNs do not reside in a significantly different environment or even systematically avoid high galaxy-density region in compar-ison with a matched control sample of galaxies (Croom et al. 2004; Sorrentino et al. 2006; Coldwell & Lambas 2006; Li et al. 2007). It appears that the difference is partly attributed to luminosity or black hole mass dependence of clustering properties and partly to the corrections of various selection effects (e.g., Croom et al. 2005; Fine et al. 2006; Myers et al. 2006; da Ângela et al. 2008; Coil et al. 2009; Hickox et al. 2009). Serber et al. (2006) studied the environments of quasars ($z \leq 0.4$, $M_i \leq -22$), reported that quasars reside in higher local overdense regions and the local density excess increases with decreasing scale at distances of less than 0.5 Mpc of quasars. They also found that there is a luminosity dependence of the density enhancement. Strand et al. (2008) explored the relationship between AGN environments and the type, luminosity, and redshift of the AGN itself, and reached a similar conclusion that higher luminosity ($M_i \leq -23.2$, $\bar{M}_i = -23.87$) AGNs have more over-dense environments compared to lower luminosity ($M_i > -23.2$, $\bar{M}_i = -22.75$) AGNs. They also presented marginal evidence for a redshift evolution of type I quasar environments and no difference between the enviroments of type I and type II quasars. These studies are consistent with the popular scenario about the formation of AGN from the simulations: quasars are triggered through gas-rich mergers while low luminosity AGNs are derived by secular evolution, which depends much weakly on the environment (e.g., Hopkins & Hernquist 2009; Lutz et al. 2010; Mullaney et al. 2010).

[1] Polar Research Institute of China, 451 Jinqiao Road, Shanghai, 200136, China; zhangshaohua@pric.gov.cn
[2] Key Laboratory for Researches in Galaxies and Cosmology, Department of Astronomy, University of Sciences and Technology of China, Chinese Academy of Sciences, Hefei, Anhui, 230026, China; twang@ustc.edu.cn,whywang@mail.ustc.edu.cn

At higher redshifts, a consensus has yet to be reached on whether quasar clustering depends on the redshift or quasar properties. Barr et al. (2003) indicated that there is no evidence for a redshift dependence of the environments from observations of 21 radio loud quasars at $0.6 < z < 1.1$. Adelberger & Steidel (2005b) reported a similar result for quasars at redshifts from 1.5 to 3.5. Coil et al. (2007, hereafter C07) found that the cross-correlation amplitude of quasars to galaxies is similar to the auto-correlation function of DEEP2 galaxies, and no significant dependence was found on either luminosity or redshift. However, Croom et al. (2002, 2005) showed the redshift-space two-point correlation amplitude has weak dependence on quasar luminosity, but increases with redshift in the range $0.5 < z < 2.5$. By combing of the 2dF quasar redshift survey (2QZ) with the fainter 2dF-SDSS LRG and QSO (2SLAQ) survey, da Ângela et al. (2008) revealed a stronger redshift dependence, and confirmed the independence of luminosity at a fixed redshift. Shirasaki et al. (2011) obtained that AGNs at higher redshift ranges reside in a denser environment than those at lower redshifts, and faint and bright AGNs displayed similar correlation amplitudes at redshifts $0.3 - 1.8$ through measuring the overdensity of galaxies around quasars using the photometric data derived from deep Subaru Suprime-Cam images.

Unlike the optically-selected type I AGNs in the above studies, the X-ray selected AGN sample, especially with Chandra and XMM, is not strongly biased agianst obscured ones. Also X-ray observations provided the deepest AGN surveys. The clustering analysis based on the spectroscopic galaxy samples and the X-ray selected AGN samples from the ROSAT ALL-Sky Survey (RASS) (e.g., Krumpe et al. 2010), Chandra and XMM data (e.g., Gilli et al. 2005, 2009; Yang et al. 2006; Coil et al. 2009) can tell us the clustering properties of the distant low luminosity AGNs or the ones missed by color-based optical surveys. Coil et al. (2009) found that the X-ray selected AGNs in red host galaxies are significantly more clustered than those in blue host galaxies, but no dependence of clustering on optical or X-ray luminosity or hardness ratio is found. However, Krumpe et al. (2010) detected a significant X-ray luminosity dependence of the clustering amplitude (see also Koutoulidis et al. 2013). The clustering amplitude for low $L_X$ AGNs is similar to that of blue star-forming galaxies, high $L_X$ sample is consistent with the clustering of red galaxies.

The controversy results may be caused by selection effects of galaxy and AGN samples. In most experiments, the galaxy samples are derived either spectroscopically or photometrically. In a spectroscopic sample, the redshifts of galaxies can be precisely measured, but only bright galaxies are targeted. Moreover, the spectroscopic targets are usually selected via optical colors, that may introduce selection bias against certain type of galaxies. For the multi-wavelength photometric sample, the galaxy samples are selected with certain color cuts and the galaxy redshifts are estimated through photometric redshift methods, which are less accurate and subject to some outliers. Thus, an unbiased galaxy sample is the most important in the investigation of the AGN environment.

In this paper we present the measurements of the quasar-galaxy cross-correlation through the projected galaxy number counts around quasars, using all photometric data from SDSS Stripe 82 at $0.6 < z < 1.2$. In this work, we do not use any color cut to select our galaxy sample. The quasar and galaxy samples used in this paper are described in Section 2. The method for calculating over-densities, the galaxy luminosity function at observers' frame and the estimate of cosmic variance are described in Section 3. The results of clustering analysis are presented in Section 4. The implication of our findings is discussed in Section 5. Throughout this paper, we assume a $\Lambda$-dominated cosmology with $H_0 = 71$ km s$^{-1}$Mpc$^{-1}$, $\Omega_M = 0.28$ and $\Omega_\Lambda = 0.72$. We define $h = H_0/(100$ km s$^{-1}$Mpc$^{-1})$ and quote correlation lengths in co-moving $h^{-1}$Mpc.

## 2. THE DATASETS AND DATA SELECTION

The SDSS imaged 11,663 deg$^2$ sky in five broad bands $u$, $g$, $r$, $i$, $z$, with the effective wavelengths of 3550, 4770, 6230, 7620 and 9130 Å, respectively (Fukugita et al. 1996; Gunn et al. 1998, 2006; Smith et al. 2002) and obtained spectra of over a million galaxies and quasars in over 8000 square degree of high Galactic latitude sky (York et al. 2000; Abazajian et al. 2009). The final photometric catalog provides accurate astrometry (Pier et al. 2003) and photometry for over 357 million unique objects (Lupton et al. 1999, 2001; Hogg et al. 2001; Smith et al. 2002; Stoughton et al. 2002; Ivezić et al. 2004; Tucker et al. 2006; Padmanabhan et al. 2008).

In SDSS imaging survey, the Celestial Equator Stripe (Stripe 82 ) is a repeating photometry rectangular region covering about 270 deg$^2$ ($\delta_{J2000}$ from -1.266 to +1.266 degree and $\alpha_{J2000}$ from -59 to 60 degree; see panel (a) of Figure 1), including approximately $12 \times 739$ contiguous fields, of size $9' \times 13'$ each. It has been imaged approx-imately 80 times (Figure 3 of Abazajian et al. 2009) and the photometry reaches $\sim 2$ magnitudes deeper than any individual scan. Due to different observing conditions, the photometric depths of these fields in Stripe 82 are not uniform and almost go deeper as the right ascension ($\alpha_{J2000}$) increases.

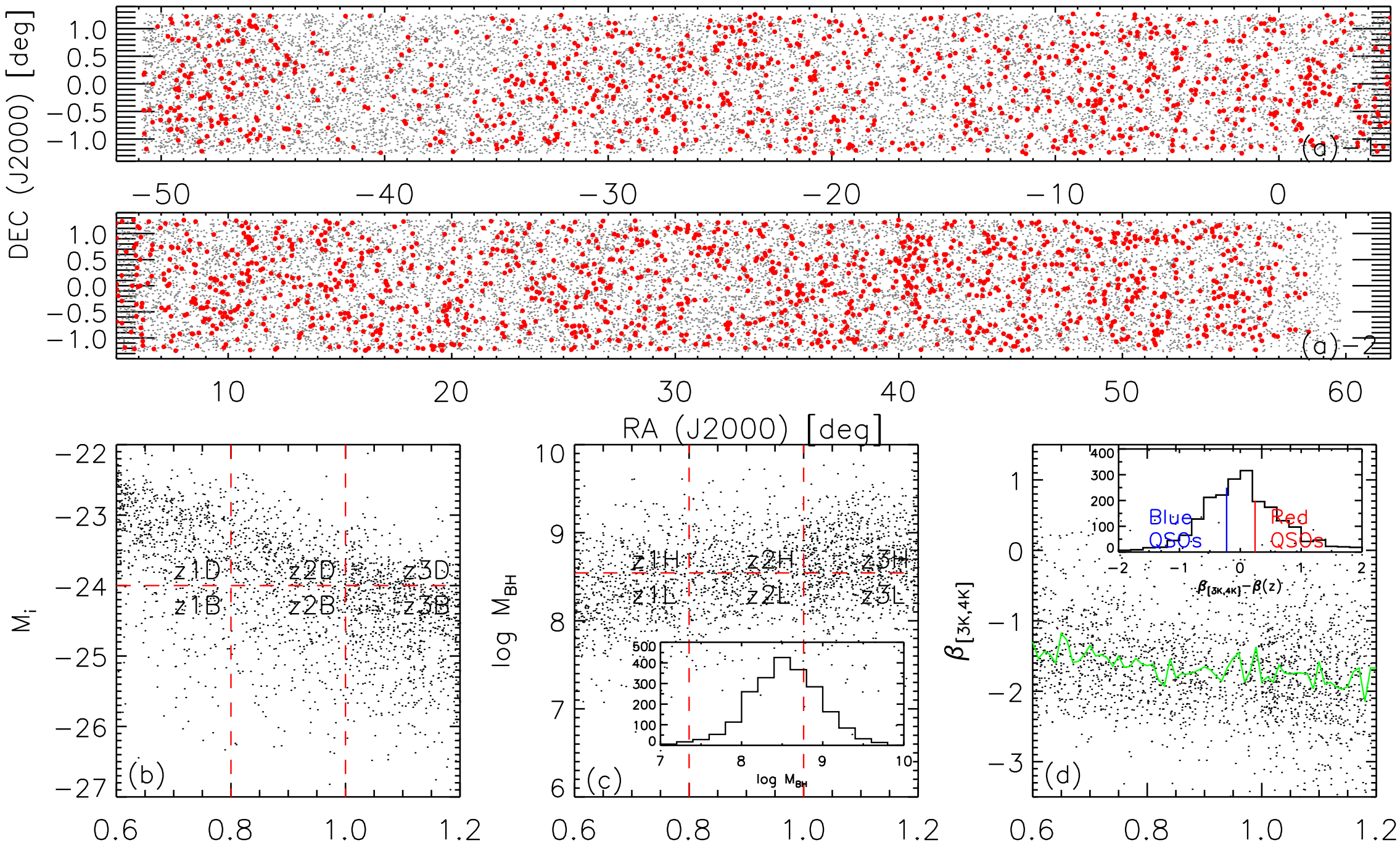

FIG. 1.— Panel (a): The distribution of photometric objects (gray dots) and quasars from Schneider et al. 2007 (red filled circles) with $0.6 < z < 1.2$ in SDSS Stripe 82 field. Panel (b, c, d): Plots of quasar luminosity ($M_i$), black hole mass ($M_{\rm BH}$) and optical color ($\beta_{[3K,4K]}$) against the redshift. The green curve in panel (d) show the median $\beta_{[3K,4K]}$ as function of redshift ($\bar{\beta}(z)$). The relative optical colors of blue quasars are $\beta_{[3K,4K]} - \bar{\beta}(z) < -0.22$ and $\beta_{[3K,4K]} - \bar{\beta}(z) > 0.24$ for the red quasars.

From the co-added images of Stripe 82 , we find the photometry precision is highest at $i$-band among the five bands [3]. To determine the survey depth of a field, we calculate the apparent magnitude distribution for all sources in the field. Since the number will drop quickly at magnitudes fainter than the survey depth, there is well-defined peak in this distribution. We choose the magnitude, which the distribution peaks at, as a magnitude threshold ($i_{\rm th}$) of the field under consideration. Assuming the galaxy number $N(i)$ at apparent magnitude $i$ follows $logN(i) \propto i$, we can extrapolate the number counts at magnitude of $i < i_{\rm th}$ to $i_{\rm th}$ to obtain a reference number for each field. We calculate the completeness by comparing the observed peak count to the reference number. The average completeness at the magnitude of $i_{\rm th}$ is 0.95, with a standard deviation of 0.06. In §4.1, we will discuss the effect of the incompleteness of the galaxy sample on our measurement. Figure 2 shows the distribution of $i_{\rm th}$ for all fields in Stripe 82 . The value of $i_{\rm th}$ has a great dispersion, from 21.2 to 23.8, with a mean value of 23.2. In this work, we select the photometric *galaxies* in Stripe 82 with $i$-band apparent magnitude less than $i_{\rm th}$ of the field.

Our *galaxy* sample is selected from the SDSS DR7 Stripe 82 calibrated object catalog (SDSS run = 100006 (South strip) and 200006 (North strip) in the DAS, 106 and 206 in the CAS; Abazajian et al. 2009), but we do not use photometric redshift information. I do not remove stars from the sample either because they will not significantly affect our results. The reasons are as follows. First, Stripe 82 is a high galactic latitude region, which is not a densely populated areas for stars. The stars constitute about 7% of all photometric objects. Second, although there is a significant gradients in the stellar density Stripe 82, its amplitudes on the scales of interest are very small. Finally, there is no spatial correlation between quasars and stars so that the stars can be considered as background using our random quasar samples (see §3.1). This is verified in §4.1 (Figure 6 and 7), in which we demonstrate that clustering signature does not present in the random fields.

We take our quasar sample from SDSS DR5 quasar catalog (Schneider et al. 2007). In the Schneider's catalog, spectroscopic targets are selected mostly accroding to their location in multidimensional SDSS color space, supplemented by X-ray and radio detected sources. Approximately half of quasars have $i < 19$, nearly all have $i < 21$. These two values correspond to the $i$ magnitude limits for $z < 3$ and $z > 3$ candidates selected by optical colors, i.e., *ugri* and *griz* color cubes (Richards et al. 2002; Schneider et al. 2007). The final catalog contains 77,429 objects, which have at least one emission line with FWHM larger than 1000 km s$^{-1}$ or interesting/complex absorption features. The redshift range is from 0.08 to 5.41. The K-corrected $i$-band absolute magnitudes of these quasars are more luminous than $M_i = -22.0$. The catalog provides a position with an accuracy better than $0.2''$ rms, five-band CCD-based photometry with typical accuracy of 0.03 mag. There are 9244 quasars fall in the Stripe 82 area. According the photometric magnitude limits and galaxy luminosity function presented in §3.2, we set the redshift range to $0.6 < z < 1.2$ to ensure that a significant fraction of galaxies are detected in the SDSS Stripe 82 . Applying this redshift cut, we obtain 2300 quasars in the final sample.

[3] $i$-band means SDSS $i$-band in this paper, and the i-band magnitude we use is the $'cmodel'$ magnitude and corrected for the Galactic extinction using the extinction map of Schlegel et al. (1998) and the reddening curve of Fitzpatrick (1999).

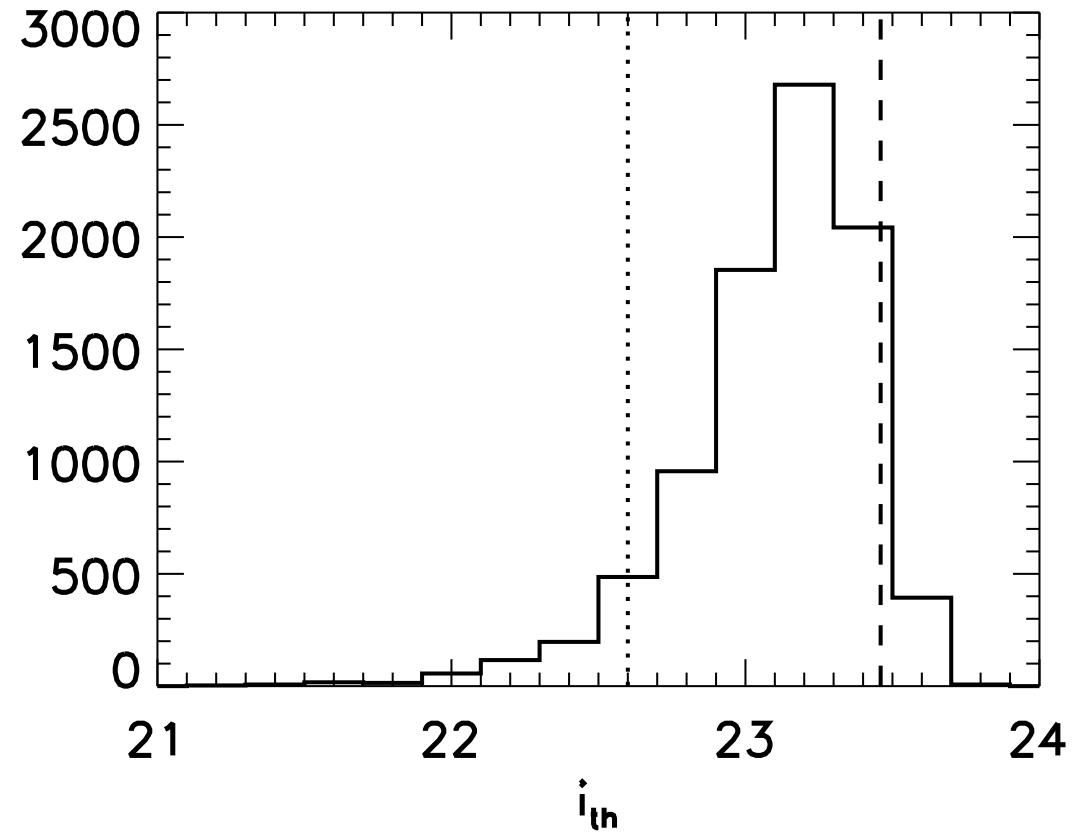

FIG. 2.— The distribution of $i_{\rm th}$ for all the fields in Stripe 82 . The value of $i_{\rm th}$ has great dispersion, from 21.2 to 23.8. The dotted line and dashed line show the 10 and 90 percentiles of the distribution.

In panel (a) of Figure 1, we show the spatial distribution of these quasars (red filled circles). There is an under-density zone at $\alpha_{J2000} \approx -40$ degree. As mentioned at the beginning of this paragraph, the quasar candidates are mainly selected from the photometric objects by their colors. We find that there are scarcely any objects with $i > 19$ at $-45 < \alpha_{J2000} < -35$ degree in the figure of $\alpha_{J2000}$-$i$ diagram for all DR5 quasars located in Stripe 82 , the so-called high-redshift candidates are absent. The only dozens are selected from X-ray and radio objects. This does not affect our final results as we are not interested in the high redshift quasars.

## 3. THE CROSS-CORRELATION OF QUASARS AND GALAXIES

### 3.1. *Galaxy Density Profile around Quasars*

The cross correlation function between quasars and galaxies, $\xi(r)$, measures the excess probability of finding a galaxy in a volume at a distance $r = (r_p^2+\pi^2)^{1/2}$ from a randomly chosen quasar, where $r_p$ and $\pi$ are coordinates perpendicular to and along the line of sight, respectively (Peebles 1980). To measure the correlation function from the spatial distribution of galaxies, one usually counts galaxy-quasar pair over a two-dimensional grid of separations to estimate $\xi(r_p, \pi)$, and then integrate $\xi(r_p, \pi)$ along $\pi$ direction up to a certain separation to eliminate the redshift distortions in $\pi$ direction (e.g. Peebles 1980). Because spectroscopic redshifts are not avialiable for galaxies in SDSS Stripe 82 region, we can not use this conventional method. In this section we will describe the projected number density method to measure the cross-correlation between quasars and galaxies.

In this method, all the photometric objects are projected on the celestial sphere. For a quasar with known redshift $z$, we calculate the projected distance ($r_p$) between a photometric object and the quasar assuming that both objects are at the same redshift, and then the projected number density of galaxies ($n(r_p)$) as a function of the projected distance to the quasar ($r_p$) are obtained. Naturally, $n(r_p)$ includes the contributions of foreground and background galaxies and stars, and the galaxies surrounding the quasar. The latter item has close connection to the cross correlation function.

By assuming a power-law form of the correlation function (e.g. Yee & Green 1987), the average number density ($\rho(r)$) of galaxies at a distance ($r$) from the quasar can be written as:

$$\rho(r) = \left[\left(\frac{r_0}{r}\right)^{\gamma} + 1\right] \rho_0, \tag{1}$$

where $r_0$ is the cross-correlation amplitude and $\gamma$ is the power-law slope, and $\rho_0$ is the galaxy density at the redshift of the quasar. The first term in the bracket represents the cross-correlation. We rewrite $\rho(r)$ as a function of $r_p$ and $\pi$ (Davis & Peebles 1983):

$$\rho(r_p, \pi) = [1 + (\pi/r_p)^2]^{-\gamma/2}(r_0/r_p)^{\gamma}\ \rho_0 + \rho_0. \tag{2}$$

Then we derive the projected galaxy number density $n(r_p)$:

$$\begin{aligned}
n(r_p) &= \int_{-\pi'}^{\pi'} d\pi\rho(r_p, \pi) + n'_{bg} \\
&= \rho_0 \left(\frac{r_0}{r_p}\right)^{\gamma} \int_{-\pi'}^{\pi'} d\pi \left[1 + \left(\frac{\pi}{r_p}\right)^2\right]^{-\gamma/2} + 2\pi' \rho_0 + n'_{bg} \\
&\doteq \rho_0 r_p \left(\frac{r_0}{r_p}\right)^{\gamma} \int_{-\infty}^{\infty} d\left(\frac{\pi}{r_p}\right) \left[1 + \left(\frac{\pi}{r_p}\right)^2\right]^{-\gamma/2} + n_{bg} \\
&\quad (if\ \pi' \gg r_p) \\
&= \rho_0 r_p \left(\frac{r_0}{r_p}\right)^{\gamma} \frac{\Gamma\left[1/2\right]\Gamma\left[(\gamma-1)/2\right]}{\Gamma\left[\gamma/2\right]} + n_{bg} \\
&= \rho_0\, \omega(r_p) + n_{bg},
\end{aligned} \tag{3}$$

where $\Gamma$ is the Gamma function. The total background density $n_{bg}$ is the sum of the density of foreground and background galaxies and stars ($n'_{bg}$), and the average galaxy density ($2\pi' \rho_0$) at the quasar's redshift. $\omega(r_p)$ is the projected cross-correlation function, which is obtained by integrating the two-point cross-correlation function $\xi(r) = \rho(r)/\rho_0 - 1 = (r_0/r)^{\gamma}$ along the line of sight (Peebles 1980). To derive the values of $r_0$ and $\gamma$, we need to know $n(r_p)$, $n_{bg}$ and $\rho_0$ from observational data.

The projected and background surface densities, $n(r_p)$ and $n_{bg}$, can be estimated using photometric catalog. Let us consider the surface density around the $j$-th quasar first. In the case that all nearby galaxies around the quasar falls in the same field of depth $i_1$, then the surface density $n_j(r_p)$ for galaxies brighter than $i_1$ is simply the number of galaxies in the annulus between radius $r_p$ and $r_p + \Delta r_p$ and the area of the annulus. When the annulus intersects with more than one field, we measure $n_j(r_p)$ using 'effective' area for each depth as we illustrate in the following example.

As showing in Figure 3, the annulus intersects with four fields (labeled as 'A' to 'D') with photometric depths in $i$-band as 22.4, 23.2, 23.6 and 23.8, respectively. If we calculate $n_j(r_p)$ within the $i$-band magnitude range $21.0 < i < 22.4$, the four fields are all *'effective'* fields, then $n_j(r_p) = N_j(r_p)/S(r_p)$, where $S(r_p)$ is the area of an intersection between the *'effective'* fields and an annulus with projected radii between $r_p$ and $r_p + \Delta r_p$. In other words, $S(r_p)$ is the sum of the green shaded area in the four fields shown in Figure 3. If the $i$-band magnitude range of interest is $23.2 < i < 23.6$, then the *'effective'* fields is just the last two fields, and $S(r_p)$ is the sum of the shaded area in the last two fields (signed as 'C' and 'D'). $N_j(r_p)$ is total number of photometric objects in the area $S(r_p)$.

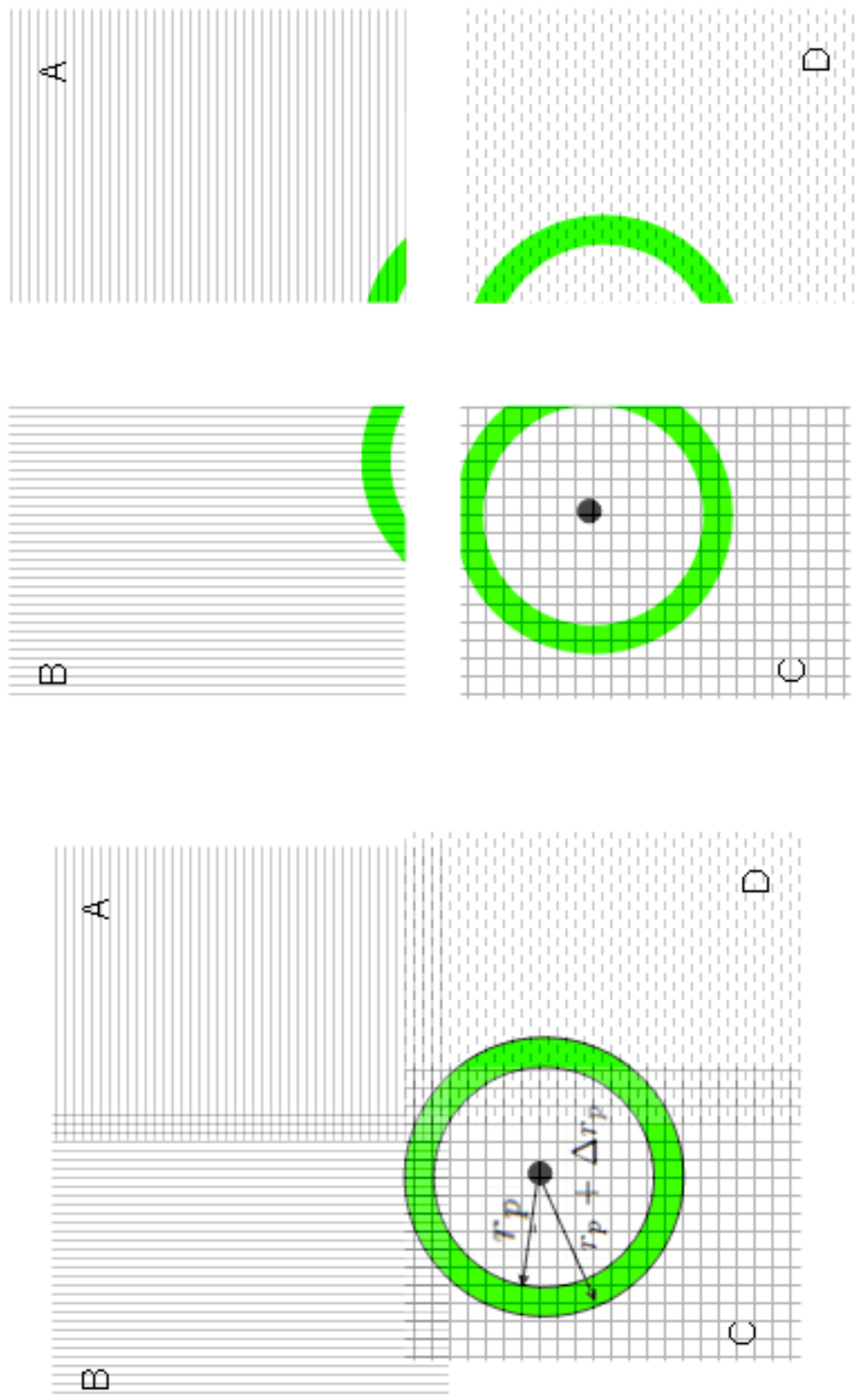

FIG. 3.— Cartoon for $n(r_p)$ measurement. Assume the photometric depth at $i$-band ($i_{\rm th}$) of the four Stripe 82 fields (signed as 'A' to 'D') are 22.4, 23.2, 23.6 and 23.8, respectively. The black filled circle is the position of a quasar. If we calculate $n(r_p)$ within the $i$-band magnitude range $21.0 < i < 22.4$, the four fields are all *'effective'* fields, then $n(r_p) = N(r_p)/S(r_p)$, where $S(r_p)$ is the area of an intersection between the *'effective'* fields and an annulus with projected radii between $r_p$ and $r_p + \Delta r_p$, is the sum of the green shaded area in the four fields. However, if the $i$-band magnitude range we want to calculate is $23.2 < i < 23.6$, then the *'effective'* fields is just the last two fields, and $S(r_p)$ is the sum of the shaded area in the last two fields (signed as 'C' and 'D'). $N(r_p)$ is total number of photometric objects in the region of $S(r_p)$.

For a total of $m$ quasars, the projected number density for each magnitude bin is estimated as: $n(r_p) = \sum_{j=0}^{m} N_j(r_p)/\sum_{j=0}^{m} S_j(r_p)$. Here $S_j(r_p)$ is the area of an intersection between the *'effective'* fields ($i_{\rm th} \geqslant i_1$) and an annulus with projected radii between $r_p$ and $r_p + \Delta r_p$ from the $j$-th quasar, and $N_j(r_p)$ is the number of galaxies in the intersection.

To estimate $n_{bg}$, we construct $N_{\rm r} = 20$ mocked random sample by placing $m$ objects randomly in the Stripe 82 field. We assign the redshift of quasars to these faked objects so that the redshift distribution in each sample is the same as our quasar sample. According to Eq. 3, the projected galaxy number density around objects is equal to $n_{bg}$, since they are not spatially correlated ($\omega(r_p) = 0$). We therefore apply the method shown above to determine $n_{bg}$ for each random quasar sample. Finally, $n_{bg}$ is the averaged value over the results for these $N_{\rm r} = 20$ random position sample. We compute $\rho_0$ as $\rho_0 = \sum_{j=0}^{m} \rho_{j,0}/m$, where $\rho_{j,0}$ is the number density of galaxies with $i$-band apparent magnitude of $i_2 < i < i_1$ at the redshift similar to the $j$-th quasar. In other words, $\rho_{j,0}$ is integration of luminosity function of galaxies from $i = i_2$ to $i_1$. The error on $\rho_{j,0}$ is propagated from the errors of parameters of the Schechter luminosity function, and then we finally obtain the error on $\rho_0$ based on the errors on $\rho_{j,0}$. The errors of Schechter function parameters are correlated, as such it will overestimate error bar. The luminosity function will be constructed in next subsection.

### 3.2. *Estimation of $\rho_0$ from Luminosity Function*

As mentioned in the last paragraph of §3.1, we need to know the number density of galaxies in an apparent magnitude range at the redshift of a quasar in question. In principle one can derive this value by integrating the galaxy luminosity function at that redshift over a luminosity range, which is transformed from the given range of apparent magnitude. The galaxy luminosity functions have been obtained by many authors at redshift $0.6-1.2$ (e.g., Gabasch et al. 2004, 2006; Ilbert et al. 2005, 2006; Zucca et al. 2006; Cirasuolo et al. 2007). These luminosity functions are all at a specific band *in the rest frame* of galaxies, while we are interested in luminosity functions at the *observed wave-band*. Because we do not know the redshift of our photometric galaxies, it is impossible to apply $k$-correction. Therefore, we have to calculate new galaxy luminosity functions based on the other deep field observation with galaxy redshift available: a deep and homogeneous $i_{\rm Subaru}$-band selected multi-waveband catalogue (Gabasch et al. 2008) from the Cosmic Evolution Survey (COSMOS), the largest survey imaging a 2 square degree equatorial field with sufficient depth (Scoville et al. 2007). It is the optimal sample for obtaining galaxy luminosity functions this work required, and the cosmic variance will be discussed in the subsequent subsection.

This catalog with a formal completeness limit of 50% for point sources of $i_{\rm Subaru} \sim 26.7$, comprises about 290,000 galaxies with observations at eight bands ($u_{\rm CFHT}$, $B_{\rm Subaru}$, $V_{\rm Subaru}$, $r_{\rm Subaru}$, $i_{\rm Subaru}$, $z_{\rm Subaru}$, $H_{\rm KPNO}$ and $K_{s,\rm KPNO}$), which are used to derive the photometric redshifts. The accuracy of the photometric redshifts is $\Delta z/(z_{spec}+1) \leq 0.035$ with only $\sim 2\%$ outliers from the comparison with spectroscopic redshifts of 162 galaxies in the redshift range $0 \lesssim z \lesssim 3$ (Bender et al. 2001; Gabasch et al. 2004). We use several galaxy spectral energy distribution (SED) templates from Coleman et al. (1980) and Kinney et al. (1996) (see the left panel of Figure 4) to fit the eight bands observation for each object in this catalogue. Then we use the best fitting SED to derive SDSS $u$-, $r$-, $i$-band magnitudes, which we are interested in.

The luminosity function of galaxies, $\phi(L)$ , is defined as $dn(L) = \phi(L)dL$, where $dn(L)$ is the number density of galaxies with luminosity in $L \pm dL/2$. Generally, the number density is computed by dividing the number of galaxies in each magnitude bin by the volume $V_{\rm bin}$ in a given redshift interval [$z_{\rm low}$, $z_{\rm high}$]. Since some faint galaxies are invisible in the whole survey volume, the $V/V_{\rm max}$ correction must be performed (Schmidt 1968). Given the magnitude limit of the observation, we can obtain the maximum redshift $z_{\rm max}$ for each object at which this object can be observed. Then this object is weighted by $V_{\rm bin}/V_{\rm max}$, where $V_{\rm max}$ is the volume enclosed between [$z_{\rm low}$, min($z_{\rm high}$, $z_{\rm max}$)]. The resultant luminosity functions in three redshift ranges is shown in Figure 5.

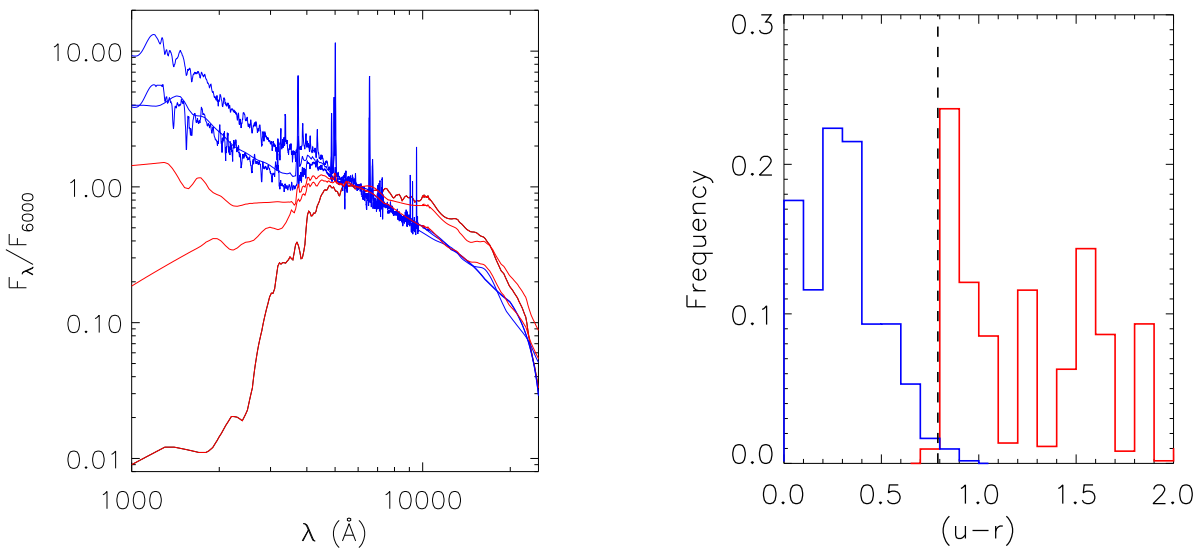

FIG. 4.— Left panel: SEDs (Empirical templates from Coleman et al. (1980) and the "SB2" and "SB3" starburst spectrum from Kinney et al. (1996)) grouped according to their spectral type. Blue curves show the blue galaxy type and red curves show the red galaxy type. More details in text. Right Panel: Color distribution between galaxies in the COSMOS field with blue SEDs (blue histogram) and red SEDs (red histogram).

The luminosity functions based on photometric redshift from Subaru might be contaminated by the spurious detections, such as stars, AGNs or blended objects in the very bright magnitude bins (see Gabasch et al. 2008). We exclude the very bright magnitude bins ($i$-band absolute magnitude $\leq -23.0$), then use the Schechter luminosity function (Schechter 1976) to fit the data. The fitting results are also shown in Figure 5 for comparison. We also derive the luminosity functions of two types of galaxies: red and blue galaxies (Figure 5), those luminosity functions will be used in the quasar and different type galaxy clustering measurements. Our test shows that the blue spectra (blue curves in the left panel of Figure 4) at the interesting redshift ($0.6 < z < 1.2$) have $u - r < 0.8$ (observed frame), while the $u - r$ color of the red spectra (red curves in the left panel of Figure 4) are redder than 0.8. Details of the test are described below: (1) SED types of galaxies with $0.6 < z < 1.2$ in Gabasch's COSMOS galaxy catalog are confirmed (the second paragraph of §3.2), (2) galaxies are divided into blue or red galaxy subsamples by their SED types, (3) photometric magnitudes at SDSS $u$- and $r$-bands are obtained from the convolution from the best fitting SED. When we compared $u - r$ color distributions of all COSMOS galaxies in blue and red subsamples, a clear line at $u - r = 0.8$ can effectively divide blue/red galaxies in the diagram (the right panel of Figure 4). [4] Thus, the dividing line is adopted as the criterion to distinguish the blue and red galaxies as we calculate the luminosity functions here, and divide the Stripe 82 galaxy sample into red and blue galaxy subsamples according to the same criterion. Actually, the criterion used in this redshift range is grounded on the SED types of galaxies, and does not concern with redshift.

[4] In fact, observed galaxy spectrum is the combination of starlights with different galactic properties and other components. Lu et al. (2006) introduced that the galaxy starlight can be sufficiently modeled with six synthesized galaxy spectral templates. COSMOS galaxies were not observed at SDSS photometric bands, the magnitudes used in the color $u-r$ are obtained from the convolution from the best fitting single galaxy template rather than the modeling results of multicomponents. There are some deviations between the magnitudes and colors we shown in Figure 4 and the direct photometric observations, and they are inclined to represent the dominant galaxy spectral template. Thus the dividing line in the $u - r$ color distribution of blue and red subsamples is more distinct than the fact.

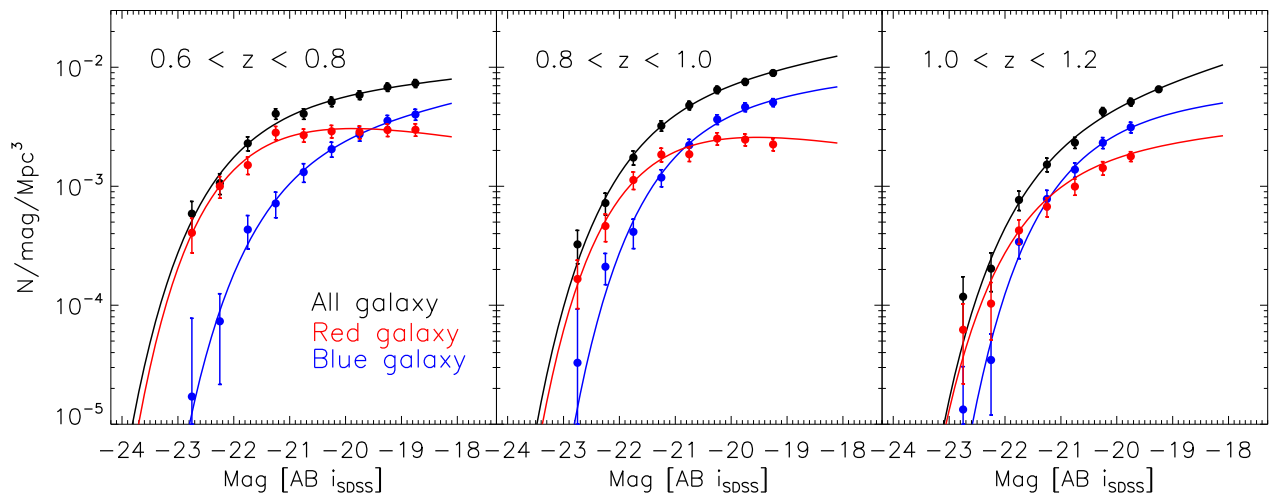

FIG. 5.— Luminosity functions in the $i$-band for the redshift intervals $0.6 < z < 0.8$ (left panels), $0.8 < z < 1.0$ (middle panels), and $1.0 < z < 1.2$ (right panels). The black lines show the best fitting Schechter functions for total luminosity functions, blue and red curves for the two SED types.

### 3.3. *Errors and Density Profile Fit*

We estimate the statistical errors of our correlation measurements using the bootstrap method (More details and different methods for realistic error calculation are referred to Norberg et al. 2009). We generate $N = 50$ bootstrap quasar samples. The objects in each bootstrap sample are randomly picked from our original quasar sample, allowing for multiple selection of the same objects. We then compute $n(r_p)$ for each sample using the method shown in §3.1. The errors of $n(r_p)$ are given by the standard deviation of the measurements among these bootstrap samples.

Adjacent radius bins in $n(r_p)$ are correlated and their errors are not independent. The covariance matrix $M_{jk}$ reflects the degree to which $j$-th radius bin is correlated with $k$-th radius bin, and has to be taken into account when we make function fits in the following. Using the number density computed from these bootstrap samples, we can derive the covariance matrix $M_{jk}$ by

$$M_{jk} = \frac{1}{N-1}\sum_{s=1}^{N}[(n(r_p^j)^s - < n(r_p^j) >) \times (n(r_p^k)^s - < n(r_p^k) >)], \quad (4)$$

where $n(r_p^j)^s$ is the galaxy number density at $j$-th radius bin for the $s$-th bootstrap sample, $< n(r_p^j) >$ is the average over all of the bootstrap samples at $j$-th radius bin (e.g., Miyaji et al. 2007). After getting the covariance matrix, we fit the density profile using Equation 3 by minimizing the correlated $\chi_c^2$ values:

$$\chi_c^2 = \sum_j^{N_{bins}}\sum_k^{N_{bins}}\left[n(r_p^j) - n^{model}(r_p^j)\right] \times M_{jk}^{-1} \times \left[n(r_p^k) - n^{model}(r_p^k)\right], \quad (5)$$

where $n(r_p^j)$ is the density profile derived from observational data while $n^{model}(r_p^j)$ is fitting curve. Then we can get the cross-correlation amplitude, $r_0$ and the slop, $\gamma$.

### 3.4. *Cosmic variance*

Cosmic variance is the uncertainty in observational estimates of the number density of galaxies in finite volumes, arising from underlying large-scale density fluctuations. It can be significant, especially for deep "pencil beam" surveys. COSMOS, used to estimate number density ($\rho_0$) of galaxies in this work, is one of examples observed at comparatively wide fields. The zCOSMOS and AzTEC/COSMOS fields, subsets of COSMOS, are known to have $\sim 3\sigma$ positive fluctuation in redshift range $z \leq 1.1$ (Meneux et al. 2009, Austermann et al. 2009). Moster et al. (2011) predicted cosmic variance of COSMOS for a given galaxy population from cold dark matter theory and the galaxy bias. In the analysis of stellar mass functions in the COSMOS field, they used stellar population synthesis models to convert luminosity into stellar mass and obtained the masses of galaxies with $i_{\rm Subaru} < 25$ (Drory et al. 2009; Ilert et al. 2010). The luminosity functions suggest that galaxies at the low-luminosity end are the dominant component in a magnitude-limited sample. The apparent magnitude of galaxies is $17.0 < i < 23.2$ in our following calculation, the absolute magnitude at $1.0 < z < 1.2$ is more luminous ($\sim 1$ mag) than that at $0.6 < z < 0.8$. In contrast with the stellar masses of COSMOS galaxies, the mean stellar masses of different galaxy samples in our magnitude range and with different redshifts, colors are in the mass range of $10^{9.2}$ M$_\odot$ $< \bar{m}_* < 10^{11}$M$_\odot$. High-redshift red galaxies have higher mass than low-redshift blue galaxies. Cosmic variance with common redshift bins gradually decrease with redshift from $\bar{z} = 0.7$ to $\bar{z} = 1.1$ with redshift bin size $\Delta z = 0.2$, and increase with galaxy stellar mass from $m_* = 10^{9.25}$M$_\odot$ to $m_* = 10^{10.75}$M$_\odot$ with mass bin size $\Delta\log(m_*/{\rm M}_\odot) = 0.25$ (Table 2 in Moster et al. 2011). The variance ranges between $\sim 8\%$ and $\sim 12\%$, the mean value is $\sim 9.5 \pm 1.1\%$. Cosmic variance can not impact the dependence of clustering on quasar properties, because the comparisons are underway in the same volume and same galaxy sample. If COSMOS is really in over-density region, the real correlation amplitudes are somewhat larger than we present in this work. However, cosmic variance is considered in the studying of dependence on galaxy properties and clustering redshift evolution. When COSMOS has $\sim 10\%$ cosmic variance, on average, the variance at $r_0$ is $\sim 5\%$, less than the measurement errors (listed in Table 2 and 3). The dependences of quasar-galaxy clustering are still present, but their confidence levels are somewhat lowered.

## 4. QUASAR-GALAXY CLUSTERING

### 4.1. *Dependence on the Redshift*

In order to investigate whether the clustering is dependent on redshift, we investigate the clustering between galaxies of $i$-band apparent magnitude in the range of (17.0, 23.2) (denoted as all galaxy sample) and quasar subsamples with different redshifts. We divide the quasar sample into three redshift intervals: 687 quasars with $0.6 < z < 0.8$ ($z1$ sample), 678 quasars with $0.8 < z < 1.0$ ( $z2$ sample ), and 935 quasars with $1.0 < z < 1.2$ ($z3$ sample). The median redshifts for these subsamples are 0.70, 0.89 and 1.09, and the median absolute magnitudes at $i$-band are -23.31, -23.87 and -24.35 mag respectively. A summary of subsample parameters can be found in Table 1. For each quasar subset, we firstly calculate $n_{bg}$ and $\rho_0$(see §3.1), which will be used as fixed values in the density profile fit procedure. We then calculate the average density profile $n(r_p)$ of galaxies as a function of the projected distance in the co-moving scales between $0.05 \leq r_p \leq 20$ $h^{-1}$Mpc from quasars as described in §3.1. The results are shown in Figure 6. As one can see, the clustering signals are detected at all redshift bins. The best fitting results ($r_0$ and $\gamma$) are listed in Table 2. For comparison, we also show the density profile for a random quasar sample at redshift of $0.6 < z < 0.8$ (see §3.1 for the method to construct random quasar sample) in the bottom-right panel of Figure 6. The density is constant with the radius, suggesting the clustering associated with the quasars is real.

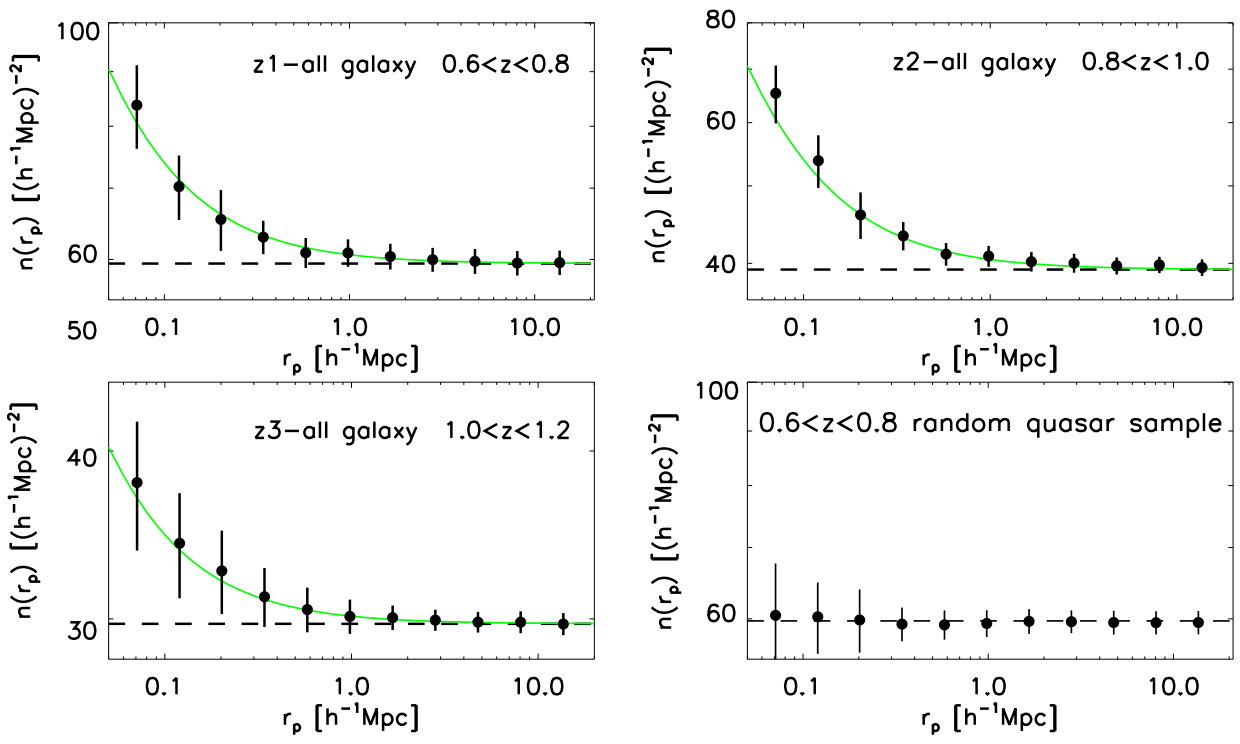

FIG. 6.— Averaged galaxy number density as a function projected distance from quasar for each redshift interval. The filled circles show the galaxy number density profile for each quasar redshift subsample, the dash lines mean the background counts ($n_{bg}$), and the solid lines show the best fitting. The right bottom panel is for random quasar sample at the redshift interval $0.6 < z < 0.8$

Initially, we fit independently the density profiles at different redshift bins with $r_0$ and $\gamma$ treated as free parameters. We find that $r_0$ increases significantly with redshift and $\gamma$ is consistent with a constant value within $1\sigma$(Table 2). To pursue better constraint on the amplitude, we fit simultaneously over the three redshift bins while locking $\gamma$ with the same value. Then we obtain the fit results: a power-law index of 2.10, and correlation amplitudes $r_0 = 3.68 \pm 0.44$ $h^{-1}$Mpc for $z1$ subsample at $0.6 < z < 0.8$; $4.91 \pm 0.40$ $h^{-1}$Mpc for $z2$ subsample at $0.8 < z < 1.0$ and $5.96 \pm 0.95$ $h^{-1}$Mpc for $z3$ subsample at $1.0 < z < 1.2$ (Table 2), consistent with free-$\gamma$ fit. The galaxies of given apparent magnitude around a high-redshift quasar is more luminous than those with the same apparent magnitude around a low-redshift counterpart. Therefore, we found the fact that the clustering of galaxies increases with the galaxy luminosity also leads to the redshift dependence. To disentangle this effect, we use galaxies in apparent magnitude ranges $17.0 < i < 22.0$ and $18.2 < i < 23.2$ to calculate the cross-correlation with $z1$ and $z3$ subsamples, respectively, the absolute magnitudes of galaxies are very similar in these two redshift bins. The resultant $r_0$ are $3.42 \pm 0.51$ $h^{-1}$Mpc and $5.77 \pm 0.92$ $h^{-1}$Mpc. Obviously,

TABLE 1
QUASAR SAMPLE

| Quasar subsample | Redshift Interval | Median $z$ | Number of Objects | Median $M_i$ | Median $M_{\rm BH}$ | Selection |
|---|---|---|---|---|---|---|
| z1 | 0.6 - 0.8 | 0.70 | 687 | -23.31 | | |
| z2 | 0.8 - 1.0 | 0.89 | 678 | -23.87 | | |
| z3 | 1.0 - 1.2 | 1.09 | 935 | -24.35 | | |
| z1D | 0.6 - 0.8 | 0.69 | 541 | -23.14 | | |
| z2D | 0.8 - 1.0 | 0.88 | 389 | -23.54 | | $M_i > -24.0$ |
| z3D | 1.0 - 1.2 | 1.06 | 243 | -23.78 | | |
| z1B | 0.6 - 0.8 | 0.73 | 146 | -24.40 | | |
| z2B | 0.8 - 1.0 | 0.91 | 289 | -24.62 | | $M_i \leq -24.0$ |
| z3B | 1.0 - 1.2 | 1.10 | 692 | -24.60 | | |
| z1L | 0.6 - 0.8 | 0.69 | 449 | -23.23 | $1.5 \times 10^8$ | |
| z2L | 0.8 - 1.0 | 0.88 | 294 | -23.79 | $1.9 \times 10^8$ | $\rm M_{BH} \leq 3.5 \times 10^8 M_\odot$ |
| z3L | 1.0 - 1.2 | 1.08 | 362 | -24.19 | $2.0 \times 10^8$ | |
| z1H | 0.6 - 0.8 | 0.71 | 212 | -23.54 | $5.9 \times 10^8$ | |
| z2H | 0.8 - 1.0 | 0.90 | 304 | -24.19 | $6.2 \times 10^8$ | $\rm M_{BH} > 3.5 \times 10^8 M_\odot$ |
| z3H | 1.0 - 1.2 | 1.09 | 559 | -24.52 | $7.4 \times 10^8$ | |
| Blue | 0.6 - 1.2 | 0.93 | 764 | -24.56 | | |
| Green | 0.6 - 1.2 | 0.94 | 769 | -24.13 | | see text for the criterion |
| Red | 0.6 - 1.2 | 0.94 | 767 | -23.97 | | |

TABLE 2
CROSS-CORRELATION RESULTS BETWEEN QUASAR SUBSAMPLES WITH DIFFERENT REDSHIFTS AND ALL GALAXY SAMPLE.

| samples | $r_0$ $h^{-1}$Mpc | $\gamma$ | $r_0(\gamma = 2.10)$ $h^{-1}$Mpc | $\rho_0$ $10^{-3}$Mpc$^{-3}$ | $n_{bg}$ $(h^{-1}\rm Mpc)^{-2}$ |
|---|---|---|---|---|---|
| z1-all galaxy | 3.49±0.89 | 2.14±0.21 | 3.68±0.44 | 9.09±0.07 | 59.47±0.15 |
| z2-all galaxy | 4.91±0.75 | 2.09±0.17 | 4.91±0.40 | 4.95±0.04 | 39.28±0.09 |
| z3-all galaxy | 6.00±1.90 | 2.10±0.25 | 5.96±0.95 | 1.06±0.01 | 29.75±0.08 |

the redshift dependence of quasar-galaxy clustering is still significant although somewhat weak.

To check the effect of galaxy sample incompleteness, we select a galaxy sample using a more conservative photometric depth ($i'_{\rm th} = i_{\rm th}$-0.2 magnitude). We then recalculate the cross-correlation functions between the three quasar subsamples and galaxies with the same apparent magnitude range, i.e. 17.0< $i$ <23.2. The results are $r_0 = 3.73 \pm 0.53$ $h^{-1}$Mpc, $5.03 \pm 0.47$ $h^{-1}$Mpc and $5.97 \pm 1.02$ $h^{-1}$Mpc with $\gamma = 2.10$ for the three subsamples. We also use this new sample to recalculate the clustering between different quasar subsamples and galaxy subsamples shown below, all new results are consistent with those presented in this paper. Therefore, our choice of photometric depth do not influence the results significantly. In addition, the maximum projected distance is fixed to 20 $h^{-1}$Mpc in this work. It is a factor of three times larger than the typical value of $r_0$. In principle, it is sufficient for the determination of correlation amplitude. To be safe, we measure $r_0$ and $\gamma$ based on the density profile with different maximum projected distance. Both $r_0$ and $\gamma$ change very little.

Because of the shallow photometric depth of Stripe 82 , we can not measure the quasar-galaxy clustering at the higher redshift. But the clustering measurement at higher redshift can be used to inspect our method. Since the galaxies at high redshift can not be detected, all the galaxies and stars in Stripe 82 calibrated object catalog are the background objects ($n_{bg}$ in the Eq. 3) and $w(r_p) = 0$. We select 683 quasars with redshift $2.0 - 2.2$ as the high redshift quasar sample, and show the projected galaxy number density ($n_{bg}$) as the function of the projected distance ($r_p$) in Figure 7. The density is constant with the radius, that confirms the reliability and robustness of our method.

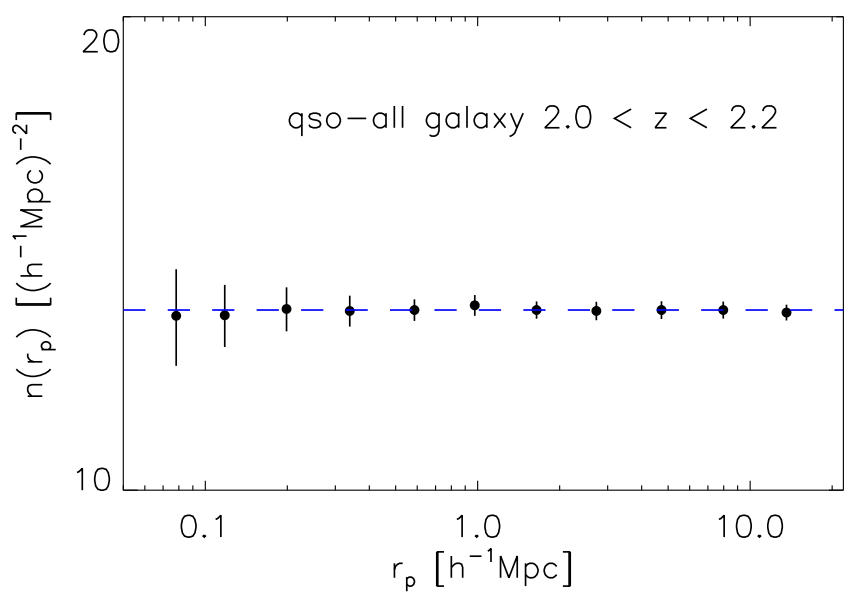

FIG. 7.— Averaged galaxy number density as a function projected distance from quasar for the redshift interval $2.0 < z < 2.2$. The dash lines is value equal to 13.02 $(h^{-1}\rm Mpc)^{-2}$.

### 4.2. *Dependence on Quasar Properties*

In this section, we try to analyze the dependence of quasar-galaxy clustering on the luminosity, black hole mass, and optical color of the quasars. The galaxy sample used here is composed of all galaxies with $i$-band apparent magnitude in the range of (17.0, 23.2). And the information for various quasar subsamples is listed in Table 1.

To examine the dependence of clustering on the luminosity of quasars, we divide all quasars of $0.6 < z < 1.2$ into luminous and faint groups with equal size. Since $\gamma$ parameter is independent of redshift, we fix it to 2.10 when fitting the density profile. We obtain $r_0 = 4.47 \pm 0.35$ $h^{-1}$Mpc for faint quasars and $5.52 \pm 0.37$ $h^{-1}$Mpc for luminous quasars. The difference is significant at $2.1\sigma$. However, this dependence can be induced by the redshift dependence of clustering because the median $z$ of faint and luminous subsamples are 0.82 and 1.04, respectively. To break the redshift-luminosity degeneracy, we split $z1$ quasar sample (the quasar sample with $0.6 < z < 0.8$) into $z1$B and $z1$D subsamples (Table 1, panel (b) of Figure 1). $z1$D subsample consists of 541 quasars in the range $-24.0 < M_i < -22.0$ and $z1$B subsample is made of 146 quasars in the range $-26.5 < M_i \leq -24.0$. The median $z$ of two subsamples are very similar. The clustering amplitudes are $r_0 = 3.76 \pm 0.40$ $h^{-1}$Mpc and $3.64 \pm 0.90$ $h^{-1}$Mpc for $z1$B and $z1$D subsamples, respectively (Table 3). Similar results are obtained for the other two redshift bins (Table 3). Therefore, we do not detect any significant dependence of quasar-galaxy clustering on quasar luminosity within our uncertainties. It is in good agreement with recent results derived using very different methods (Croom et al. 2005; Adelberger & Steidel 2005b; Myers et al. 2006; Shirasaki et al. 2011).

Next, we study the dependence of clustering on black hole mass of quasars. The black hole masses for most quasars are calculated using black hole mass formalism based on broad Mg II line and monochromatic luminosity at 3000Å. For a small number of quasars that Mg II spectral regime is not available, the masses are estimated using the broad H$\beta$ line and 5100Å monochromatic luminosity. Both Mg II- and H$\beta$-based $M_{\rm BH}$ estimators come from Wang et al. (2009). We take the median $M_{\rm BH}$ value of the whole sample, $3.5 \times 10^8$ $M_\odot$, as the arbitrary value $M_{th}$, and divide the $z$j(j=1,2,3) quasar sample into two subsamples $z$jH and $z$jL, each of them consists of quasars with $M_{\rm BH} > M_{th}$ and $M_{\rm BH} < M_{th}$ (see panel (c) of Figure 1). As shown in Table 3, $r_0$ appears larger for massive black hole mass in each redshift bin with a significance of $\sim 1\sigma$.

Finally, we examine the dependence of clustering on the optical color of quasars. The continuum slope $\beta_{[3K,4K]}$ ($f_\lambda \propto \lambda^{\beta_{[3K,4K]}}$) is measured from the Galactic reddening corrected quasar spectrum between $\sim$ 3000 Å and $\sim$ 4000 Å in the rest-frame. The two continuum windows, [3010, 3040]Å and [4210, 4332]Å, are chosen to avoid strong Fe II multiplets shortward of 3000 Å, and possible star-light contribution longward of 5000 Å in some quasars. Thus, $\beta_{[3K,4K]}$ is used as a fair measurement continuum slope. We calculate $\bar{\beta}(z)$, the median $\beta_{[3K,4K]}$ as function of redshift, and divide quasars into 'Blue', 'Green' and 'Red' quasar subsamples according to their relative optical color ($\beta_{[3K,4K]} - \bar{\beta}(z)$) (see panel (d) of Figure 1). We select different optical color subsamples in this way to make sure that these subsamples have similar redshift distribution. A summary of these subsamples is listed in Table 1. As shown in Table 3, the correlation amplitude is marginally larger ($\sim 2.3\sigma$) for blue quasars than that for red quasars. The blue quasars are slightly more luminous than the red quasars( Table 1). Because we do not find significant evidence for the dependence of clustering on quasar luminosity, the color dependence of quasar-galaxy clustering is not ascribed to the difference in luminosity. We cross-match our quasar sample with the objects in the GALEX (Morrissey et al. 2007) and UKIDSS (Lawrence et al. 2007) observation within a 2″position offset, and find 1211 and 1869 quasars, respectively. We then calculate their NUV-Optical and Optical-IR color. The clustering measurements show that the color dependence can also be detected based on color in other wavebands.

TABLE 3
Cross-correlation results for the various quasar and galaxy subsamples

| samples | $r_0(\gamma = 2.10)$ $h^{-1}$Mpc | $\rho_0$ $10^{-3}$Mpc$^{-3}$ | $n_{bg}$ $(h^{-1}Mpc)^{-2}$ |
|---|---|---|---|
| z1B-all galaxy | 3.64±0.90 | 8.57±0.12 | 55.79±0.23 |
| z2B-all galaxy | 4.86±0.58 | 4.76±0.06 | 39.28±0.14 |
| z3B-all galaxy | 6.12±1.03 | 1.03±0.01 | 29.63±0.09 |
| z1D-all galaxy | 3.76±0.45 | 9.23±0.08 | 60.17±0.17 |
| z2D-all galaxy | 4.93±0.49 | 5.11±0.05 | 39.32±0.13 |
| z3D-all galaxy | 6.06±1.13 | 1.17±0.02 | 29.75±0.10 |
| z1H-all galaxy | 4.50±0.56 | 8.86±0.08 | 56.82±0.25 |
| z1H-all galaxy | 5.13±0.51 | 4.91±0.06 | 39.73±0.13 |
| z1H-all galaxy | 6.77±1.17 | 1.04±0.02 | 29.44±0.09 |
| z1L-all galaxy | 3.51±0.49 | 9.17±0.12 | 60.05±0.16 |
| z1L-all galaxy | 4.43±0.55 | 5.06±0.06 | 38.86±0.15 |
| z1L-all galaxy | 5.52±1.24 | 1.09±0.01 | 29.96±0.09 |
| Blue-all galaxy | 5.57±0.37 | 4.59±0.04 | 35.59±0.14 |
| Green-all galaxy | 4.53±0.45 | 4.57±0.04 | 35.58±0.15 |
| Red-all galaxy | 4.02±0.59 | 4.62±0.04 | 36.55±0.13 |
| z1-Bright Galaxy[a] | 6.73±0.62 | 0.83±0.01 | 17.91±0.21 |
| z2-Bright Galaxy[b] | 7.84±0.84 | 0.28±0.03 | 15.09±0.16 |
| z3-Bright Galaxy[c] | 16.52±2.97 | 0.06±0.00 | 15.98±0.12 |
| z1-Faint Galaxy[d] | 3.71±0.52 | 8.27±0.11 | 41.57±0.14 |
| z2-Faint Galaxy[e] | 4.55±0.46 | 4.69±0.05 | 24.18±0.11 |
| z3-Faint Galaxy[f] | 5.78±1.04 | 1.00±0.01 | 13.78±0.10 |
| z1-Blue Galaxy[g] | 4.12±0.36 | 2.52±0.06 | 10.11±0.07 |
| z2-Blue Galaxy | 5.24±0.39 | 1.79±0.03 | 6.60±0.05 |
| z3-Blue Galaxy | 6.62±0.65 | 0.59±0.01 | 5.02±0.04 |
| z1-Red Galaxy | 3.64±0.62 | 5.96±0.07 | 49.36±0.17 |
| z2-Red Galaxy | 3.89±0.86 | 2.60±0.03 | 32.69±0.11 |
| z3-Red Galaxy | 4.63±1.51 | 0.47±0.01 | 24.71±0.08 |

Note. — The apparent magnitude ranges of the galaxy samples are a: $17.0 < i < 21.0$, b: $17.0 < i < 21.4$, c: $17.0 < i < 22.0$, d: $21.0 < i < 23.2$, e: $21.4 < i < 23.2$, and f: $22.0 < i < 23.2$. Please see the text (§4.3) for why we select galaxy samples in this way. g: the criterion for red and blue galaxy samples is $u - r = 0.8$ (§3.2).

### 4.3. *Dependence on Galaxy Properties*

The galaxy clustering depends on galaxy properties, such as color (e.g., Zehavi et al. 2002, 2005; Coil et al. 2004; Li et al. 2006), and luminosity (e.g., Zehavi et al. 2002, 2005; Li et al. 2006), in the local or intermediate redshift ($0.2 < z < 1$) universe. The red/bright galaxies are much more strongly clustered than blue/faint galaxies. The cross-correlations between quasars and galaxies also rely on the galaxy properties. In this section, we will use $z1$, $z2$ and $z3$ quasar samples to examine how the correlation amplitude changes with the galaxy luminosity or spectral types. These will provide additional constraints on galaxy and AGN co-evolution model. Here again, we fix the slop to $\gamma = 2.10$ during the fitting procedure.

Firstly, in order to investigate the dependence on galaxy magnitude, we divide the galaxies into two apparent magnitude bins. The magnitude threshold used to split the galaxy sample changes with the cross-correlated quasar sample. For example, as we calculate the density profile of bright and faint galaxies around $z1$ quasar sample, the apparent magnitude threshold is adopted as $i = 21.0$. While we calculate the density around $z3$ sample, the magnitude threshold increases to 22.0(see Table 3). The magnitude threshold is so chosen that the corresponding absolute magnitude is approximately same after taking into consideration average $K$-correction at each redshift bin, the values of $K$-correction are roughly estimated through the mean galaxy SED. All the results are shown in Table 3. The correlation amplitudes with faint galaxies are very similar to those with all galaxies (Table 1), because the number of faint galaxies are much more abundant than luminous counterparts. In all three redshift bins, correlation amplitudes with luminous galaxies are about two times larger than those with faint galaxies. The differences are significant at 3.7, 3.6 and $3.8\sigma$ levels. In addition, $r_0$ significantly increases with redshift, especially for luminous galaxies, consistent with our previous results.

We further explore how quasar-galaxy clustering depends on the galaxy SED. We take $u-r = 0.8$ as the color criterion and divide the photometric galaxies into red and blue galaxy samples (see §3.2 for reasons). We then measure the cross-correlation between $z$j quasar samples and these two galaxy samples ($17.0 < i < 23.2$). The results are shown in Table 3. For blue galaxies, the correlation amplitude increases remarkably with redshift (Figure 8). However, the correlation amplitude for red galaxies varies only marginally. And there is a significant trend that the correlation amplitude is larger for blue galaxies than for red galaxies. The small scale slope of the correlation function is known to vary with galaxy color, quasar-blue galaxy clustering has a slightly steeper slope than that of quasar-red galaxy (e.g., Norman et al. 2009), the fixed gamma extends the difference on the amplitude between the clustering of quasar with both type galaxies. quasar-blue galaxy clustering shows a greater amplitude and quasar-red galaxy clustering shows a smaller amplitude than those of we fit using free slopes.

### 4.4. *Comparisons with other results*

Generally, the different clustering measurements can not be compared directly unless the studies used the same luminosity/redshift range of AGNs/quasars and the same luminosity/redshift range of galaxies. Comparing to this work, (most of) other AGN samples have much lower luminosity and (most of) other galaxy samples have much higher luminosity in the measurement of AGN-galaxy CCFs. However, the variation trend of the clustering with the AGN/quasar or galaxy property changing obtained in other studies can be used to check up the clustering dependence obtained in this work. Furthermore, the comparison between our work and other studies also tell us the potential reason for the difference results and the non-dependence on the AGN/quasar and galaxy properties for the similar results.

At redshift $z = 0.6 \sim 0.8$, cross-correlation amplitude derived in this paper is consistent with these derived from cross-correlation of infrared-selected and X-ray selected AGNs in AGES with spectroscopic galaxy sample (Hickox et al. 2009) over a smaller area, from cross-correlation of SUBARU photometric galaxies with faint AGN sample (Shirasaki et al. 2011) over a broader redshift range, and from the analysis of X-ray AGN in Chandra DEEP North (Gilli et al. 2005). However, these amplitudes are significantly smaller than recent results based on the cross-correlation of quasars with spectroscopic of Luminous Red Galaxy (LRG; Norman et al. 2009, hereafter N09; Padmanabhan et al. 2009; Mountrichas et al. 2009) and AGN-galaxy correlation in the Chandra Deep South field (Gilli et al. 2005). In addition, these authors derived a smaller power-law index $\gamma \sim 1.65 - 1.83$.

These differences may be caused by three factors: different galaxy luminosity are involved, different weights are employed towards the large and small scales and cosmic variance. Firstly, LRG galaxies in these samples are more luminous and massive than our galaxy samples. A large clustering amplitude seems consistent with our results that luminous galaxies tend to be strongly clustered around quasars. In fact, their clustering amplitudes are similar to those of the luminous galaxies presented in this work. Secondly, our method is more sensitive to the small scale over-density than these using the spectroscopic galaxy sample because fluctuation of a large number of background/foreground sources will overwhelm the signal in the outer region. In addition, Meneux et al. (2009) showed that flux limit will tend to weaken the clustering measurement at small scale because of mass incompleteness in a flux limited sample. Since our photometric sample is deeper than these LRG spectroscopic sample, the effect of mass incompleteness tends to be smaller. As illustrated by McBride et al. (2011) in the LasDamas beta mock sample for local universe, $w(r_p)$ has a steeper slope of around 2.06 on the scales of 0.2 to 1.0 $h^{-1}$Mpc than on scales of 2-10 $h^{-1}$Mpc ($\simeq$1.70). Thus, most of the quasar-galaxy clustering signal we measure probably comes from the scales smaller than 1 $h^{-1}$Mpc, although we measure up to 20 $h^{-1}$Mpc. In other words, we probably only measure the clustering in the 1-halo term. Thirdly, the large difference in the two Chandra deep fields are likely due to cosmic variance (Meneux et al. 2009), although different depth may also explain some of the difference.

At redshift $z \sim 1$, the clustering amplitude for the whole quasar sample in Stripe 82 is significantly larger than these obtained by C07 for optically-selected quasars in AGES and galaxies in DEEP2 survey, smaller than the AGN sample of Shirasaki et al. (2011), but similar to the X-ray selected AGNs in AGES (Coil et al. 2009). C07 measured $r_0 = [2.95, 3.56]$ $h^{-1}$Mpc and $\gamma = 1.83$ over distance scales from 0.05 to 10 $h^{-1}$Mpc by cross-correlating DEEP2 galaxies at $0.7 < z < 1.4$ with different quasar sample identified in the SDSS and DEEP2 surveys. Shirasaki et al. (2011) got a very similar correlation amplitude between bright quasars and Subaru deep photometric data. When $\gamma$ is fiexed to 1.83, we will obtain a much higher correlation amplitude of $5.31 \pm 0.49$ $h^{-1}$Mpc and $6.57 \pm 1.04$ $h^{-1}$Mpc. Because the luminosity ranges of both quasars ($M_B < -22$) and galaxies ($R_{AB} < 24.2$) are similar to our samples, the reason for the discrepancy is not easily understood, it's likely to be the systematics of different methods. Bornancini et al. (2007) found a larger correlation amplitude ($r_0 = 5.4 \pm 1.6$ $h^{-1}$Mpc and $\gamma = 1.94 \pm 0.10$) in the DEEP2 fields. Their amplitude is similar to the correlation of red galaxies with quasars in our sample. Because Bornancini et al. (2007) used the distant red galaxies at redshifts from 1 to 2 with the color cut $J - K_s > 2.3$, their results are fully consistent with ours.

Clustering amplitude increasing with black hole mass but not quasar luminosity in the redshift range is considered in broad consistence with previous results. Hickox et al. (2009) analyzed the cross correlation between AGNs and galaxies based on AGN and Galaxy Evolution Survey (AGES) at redshift from 0.25 to 0.8. AGNs selected with radio, X-ray detection and infrared colors show different properties, and the radio and infrared selected AGNs have similar black hole mass to our High-$M_{BH}$ and Low-$M_{BH}$ samples, respectively. The clustering amplitude of the radio-selected AGN is significantly larger than that of infrared selected AGN, indicating an increasing clustering amplitude with the black hole mass. Coil et al. (2009) compared the clustering amplitudes of X-ray AGNs and quasars in C07, found that X-ray AGNs at redshift $z \sim 1$ are more clustered than quasars with a $2.6\sigma$ significance, meanwhile the X-ray AGNs have higher black hole mass than quasars. Summary of above mentioned, the clustering dependence on the black hole mass is confirmed in other works. Admittedly, the AGN samples in these works have significantly different properties expect the black hole mass, such as radio property, IR, optical or X-ray selection, absorbed or unabsorbed, and broad or narrow emission-lines. We also can not rule out the possibility of the increasing CCF clustering with black hole mass due to the selection or other properties (non- $M_{BH}$) of AGN samples.

Previous studies have shown that quasar auto-correlation function (ACF) and AGN-galaxy CCF clustering amplitude increases with redshift. Croom et al. (2005) divided over 20,000 2QZ objects into 10 redshift intervals with effective redshifts from $z = 0.53$ to 2.48. They measured the redshift-space two-point correlation functions, and found the quasar clustering amplitude increases with redshift such that the integrated correlation function $\bar{\xi}(z = 0.53) = 0.26 \pm 0.08$ and $\bar{\xi}(z = 2.48) = 0.70 \pm 0.17$. Myers et al. (2006) also found increasing clustering amplitude with redshift by calculating the projected angular clustering of $\sim 80{,}000$ photometric quasars in SDSS. Directly related to this work, Hopkins et al. (2007) confirmed the conclusion from comparing observed clustering of quasars and galaxies as a function of redshift (dotted line in Figure 8). Our measurement from cross-correlation over a much smaller redshift range and on smaller scales shows the same trend.

One of the most surprising finding is that cross-correlation amplitude depends strongly on the galaxy color. C07 found that quasars reside in regions of similar mean over-density to blue galaxies than red galaxies when they looked at the correlation function of the SDSS quasars with the DEEP2 galaxies at $z \sim 1$. N09 separated the 2SLAQ LRG sample which contains near 15,000 LRGs with magnitudes more luminous than $i = 19.8$ and the median redshift $\sim 0.52$ into two populations of blue and red galaxies by $g - r = 1.6$. They found the projected two point correlation to have fitted clustering amplitude of $r_0 = 7.3 \pm 0.7\ h^{-1}$Mpc and $r_0 = 4.9 \pm 0.7\ h^{-1}$Mpc on scales from $0.7 - 27\ h^{-1}$Mpc for the two populations. Those quasars have a stronger correlation amplitude with the blue population than the red population. Although the color selected methods of ours and N09 are different, the quasar correlation amplitude trend with blue and red galaxies are consistent. Furthermore, we found that the correlation amplitude with blue galaxies increases strongly with redshifts while that with red galaxies varies marginally. Note that the redshift distributions of blue and red galaxies can affect the clustering dependence on the galaxy color. However, the galaxy sample we used is Stripe 82 photometric sample, the redshifts of these galaxies are absent.

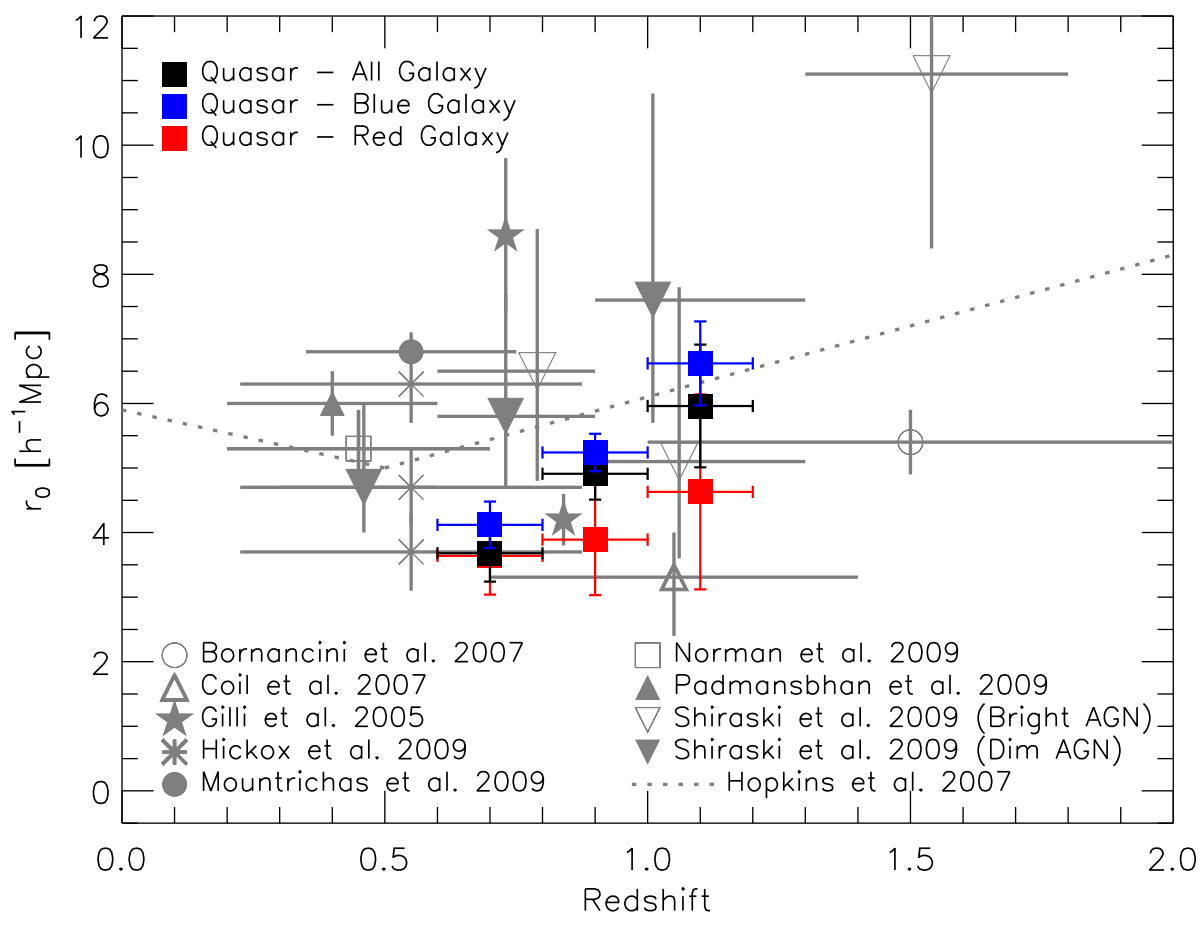

FIG. 8.— The cross-correlation amplitude as the function of redshift.

## 5. DISCUSSION

In the local universe, the clustering of galaxies measured by many authors are $5 - 7\ h^{-1}$Mpc (Hawkins et al. 2003; Zehavi et al. 2005; Li et al. 2007; Ma et al. 2009). Measurements up to the redshift $z = 2$ suggest that clustering length increases with redshift. In the redshift range of this paper, the clustering amplitudes are around $5 - 9\ h^{-1}$Mpc, depending on the mass of the galaxies (Pollo et al. 2006; Coil et al. 2006; Meneux et al. 2008; Magliocchetti et al. 2008; Foucaud et al. 2010). The clustering amplitudes of galaxies around quasars derived in this paper are similar to these values, and also with the same trend of dependence on the redshift. That implies that these quasars are located in a dark matter halo which has similar mass to that occupied by typical galaxies in these samples with stellar mass larger than $10^{10}\ M_\odot$.

Evidences have been mounted for the strong link between the black holes and the mass of the bulge components of galaxies (e.g., Maggorian et al. 1998). In Halo Occupation Distribution (HOD) models, the masses of these bulges are correlated with the mass of the halo (e.g., Jing et al. 1998; Yang et al. 2003). Therefore, quasars with higher $M_{BH}$ are located in relatively larger host dark matter halos. Indeed, Ferrarese (2002) found that the mass of the central black hole is plausibly correlated with that of its host halo (the total gravitational mass of its host galaxy). Furthermore, it is well known that the halo clustering is dependent on halo mass (e.g. Mo & White 1996). Therefore, it is expected that clustering around more massive quasars is stronger than that around lower mass quasars. Our finding is consistent with this.

The quasar luminosity is determined by the black hole mass and Eddington ratio. Thus for a given Eddington ratio, the luminosity is proportional to the black hole mass. If all quasars are accreting at the similar Eddington ratio or close to the Eddington limit, we will observe the significant correlation of clustering strength with the quasar luminosity. If the luminosity of the quasar sample is driven by largely by Eddington ratios, the luminosity dependence will be quite weak, or even disappear. Hopkins et al. (2007) discussed the connection between the observed quasar to galaxy cross correlation and luminosity. They also considered that the variation in Eddington ratios at a given black hole mass is major driver of the weak dependence on quasar luminosity, and the clustering is much more strongly correlated with galaxy luminosity than that with quasar luminosity (also see Coil et al. 2009).

We also found the quasar-galaxy clustering depends significantly on quasar's color, in the sense that blue quasars are more strongly clustered than red quasars. It strongly implies that the AGN activity is influenced by large scale environment, although the underlying physics is unclear. We also noticed that Hickox et al. (2001) measured the spatial clustering of luminous mid-infrared selected obscured and unobscured quasars in the redshift range $0.7 < z < 1.8$. Their results indicate that the cross-correlation of the obscured quasars with galaxies is somewhat stronger than that for the unobscured quasars. Generally, the obscured quasars are possible more redder than the unobscured quasars. The measurements of Hickox et al. (2011) don't agree with this work. However, it must be point that the selection boundary of the obscured and unobscured quasars is the optical-IR color color selection at $R-[4.5\mu m]_{IRAC\ band} = 6.1$ (please see Hickox et al. (2007; 2011) for more details about the mid-infrared selected quasar sample). We also estimated the similar optical-IR colors for our quasars. The magnitudes at $R$ band are transformed by the transformations from Jester et al. (2005), and IR band is instead of the W2-band observations of Wide-field Infrared Survey Explorer (WISE; Lonsdale et al. 2003). We find that all of our quasars locate at the region of unobscured quasars in the panel (b) of Figure 1 of Hickox et al. (2011). The color of quasars is a complicated parameter associated with other parameters, such as the black hole mass, Eddington ratio, and large scale environment. It needs more works to understand the underlying reason of the clustering dependence on the quasar colors.

However our finding that the clustering of blue galaxies around quasars is stronger than that of red galaxies is inconsistent with this scenario if the color dependence of galaxy clustering found in the local universe also holds at high redshift. In the local universe, blue galaxies exhibit a lower correlation amplitude, but bright red galaxies exhibit the strongest clustering at large scales while that of faint red galaxies at small scales (e.g., Zehavi et al. 2005; Loh et al. 2010). The fit to the projected correlation function with HOD models indicates that blue galaxies are field galaxies located at the central of low mass halos, the majority of less luminous red galaxies are satellites of massive halos and most luminous red galaxies are central galaxies of massive halos (Zehavi et al. 2005). This suggests that at redshift 0.6 to 1.2, quasars do not reside in massive cluster in general, but rather in less massive group of galaxies. Guo & White (2008) studied the growth of galaxies in the De Lucia & Blaizot (2007) model for galaxy formation through major and minor mergers. For low mass galaxies, star-formation and minor mergers are the dominant modes for the galaxy growth. Major mergers are much more important for significantly more massive than the Milky Way galaxies. The minor merger of low mass halos can provide a great deal of gas for star-formation and AGN accretion and ignite AGN activity. The stronger correlation of quasars with blue galaxies than with red galaxies implies minor mergers may play important role in triggering nuclear activity.

## 6. CONCLUSION

We investigate the spatial clustering of galaxies around quasars at redshifts from 0.6 to 1.2 using the photometric data from SDSS Stripe 82 . The quasar and galaxy cross-correlation functions are measured through the projected galaxy number density $n(r_p)$ on scales $0.05-20\ h^{-1}$Mpc for a sample of 2300 quasars from Schneider et al. (2007). The main results of this paper can be summarized as follows:

(1) The average clustering amplitude increases with redshifts, and blue and luminous galaxies contributes to the most of such increment, while the power-law slope of density distribution is broadly consistent with a constant value. The clustering amplitudes are $3.49\pm1.23\ h^{-1}$Mpc with $\gamma = 2.14 \pm 0.28$, $4.91 \pm 1.19\ h^{-1}$Mpc with $\gamma = 2.09\pm0.17$, and $6.00\pm1.97\ h^{-1}$Mpc with $\gamma = 2.10\pm0.35$ at redshift $z = 0.6-0.8$, $0.8-1.0$, and $1.0-1.2$. When the slope is fixed to 2.10, the clustering amplitude increases with redshifts at $2.3\sigma$ levels.

(2) The clustering amplitude varies with black hole mass, quasar's color. When we split the quasars into two black hole mass subsamples by $M_{\rm BH} = 3.5\times10^8\ {\rm M}_\odot$ at each redshift bin. The clustering amplitude is slightly larger for quasars with more massive black hole mass in each redshift interval. We also find the clustering amplitude depend on the color of quasar, the amplitude is significantly larger ($2.3\sigma$) for blue quasars than that for red quasars. While the dependence on quasar luminosity is absent in each redshift bin.

(3) There is strong dependence of clustering amplitudes on the SED type of galaxies, with blue galaxies more strongly clustered around quasars than red galaxies at $3.4\sigma$ confidence level.

Many thanks to thank Dr. Ting Xiao and Dr. Xueguang Zhang for very helpful discussion. We thank the anonymous referee for constructive comments and suggestions. This work is supported by NSFC (11233002, 11073017, 11033007), 973 project (2007CB815403) and the Fundamental Research Funds for the Central Universities and Chinese Universities Scientific Fund.

## REFERENCES

Abazajian, K. N., et al. 2009, ApJS, 182, 543
Adelman-McCarthy, J. K., et al. 2007, ApJS, 172, 634
Adelberger, K. L., & Steidel, C. C. 2005b, ApJ, 630, 50
Austermann, J. E., Aretxaga, I., Hughes, D. H., et al. 2009, MNRAS, 393, 1573
Bahcall, J. N. 1969, ApJ, 158, L87

Barr, J. M., Bremer, M. N., Baker, J. C., & Lehnert, M. D. 2003, MNRAS, 346, 229
Becker, R. H., White, R. L., & Helfand, D. J. 1995, ApJ, 450, 559
Bender R. et al. 2001, in Christiani S., ed., ESO/ECF/STScI Workshop on Deep Fields. Springer-Verlag, Berlin , p. 96
Berlind, A. A., & Weinberg, D. H. 2002, ApJ, 575, 587
Bornancini, C. G., & García Lambas, D. 2007, MNRAS, 377, 179
Brammer, G. B., van Dokkum, P. G., & Coppi, P. 2008, ApJ, 686, 1503
Bundy, K., et al. 2006, ApJ, 651, 120
Cirasuolo, M., et al. 2007, MNRAS, 380, 585
Coil, A. L., et al. 2004, ApJ, 609, 525
Coil, A. L., Hennawi, J. F., Newman, J. A., Cooper, M. C., & Davis, M. 2007, ApJ, 654, 115
Coil, A. L., et al. 2009, ApJ, 701, 1484
Coleman, G. D., Wu, C.-C., & Weedman, D. W. 1980, ApJS, 43, 393
Coldwell, G. V., & Lambas, D. G. 2006, MNRAS, 371, 786
Collister, A. A., & Lahav, O. 2004, PASP, 116, 34
Croom, S. M., Boyle, B. J., Loaring, N. S., Miller, L., Outram, P. J., Shanks, T., & Smith, R. J. 2002, MNRAS, 335, 459
Croom, S., et al. 2004, AGN Physics with the Sloan Digital Sky Survey, 311, 457
Croom, S. M., et al. 2005, MNRAS, 356, 415
da Ângela, J., Shanks, T., Croom, S. M., et al. 2008, MNRAS, 383, 565
Davis, M., & Peebles, P. J. E. 1983, ApJ, 267, 465
De Lucia, G., & Blaizot, J. 2007, MNRAS, 375, 2
De Robertis, M. M., Yee, H. K. C., & Hayhoe, K. 1998, ApJ, 496, 93
Drory, N., Bundy, K., Leauthaud, A., et al. 2009, ApJ, 707, 1595
Ellingson, E., Yee, H. K. C., & Green, R. F. 1991, ApJ, 371, 49
Ferrarese, L., & Merritt, D. 2000, ApJ, 539, L9
Ferrarese, L. 2002, ApJ, 578, 90
Fine, S., Croom, S. M., Miller, L., et al. 2006, MNRAS, 373, 613
Foucaud, S., Conselice, C. J., Hartley, W. G., Lane, K. P., Bamford, S. P., Almaini, O., & Bundy, K. 2010, MNRAS, 406, 147
Fukugita, M., Ichikawa, T., Gunn, J. E., Doi, M., Shimasaku, K., & Schneider, D. P. 1996, AJ, 111, 1748
Gabasch, A., et al. 2004, A&A, 421, 41
Gabasch, A., et al. 2006, A&A, 448, 101
Gabasch, A., Goranova, Y., Hopp, U., Noll, S., & Pannella, M. 2008, MNRAS, 383, 1319
Gebhardt, K., et al. 2000, ApJ, 539, L13
Gilli, R., et al. 2005, A&A, 430, 811
Gilli, R., et al. 2009, A&A, 494, 33
Graham, A. W. 2007, MNRAS, 379, 711
Gunn, J. E., et al. 1998, AJ, 116, 3040
Gunn, J. E., et al. 2006, AJ, 131, 2332
Guo, Q., & White, S. D. M. 2008, MNRAS, 384, 2
Hawkins, E., et al. 2003, MNRAS, 346, 78
Hickox, R. C., Jones, C., Forman, W. R., et al. 2007, ApJ, 671, 1365
Hickox, R. C., et al. 2009, ApJ, 696, 891
Hickox, R. C., Myers, A. D., Brodwin, M., et al. 2011, ApJ, 731, 117
Hill, G. J., & Lilly, S. J. 1991, ApJ, 367, 1
Hogg, D. W., Finkbeiner, D. P., Schlegel, D. J., & Gunn, J. E. 2001, AJ, 122, 2129
Hopkins, P. F., Lidz, A., Hernquist, L., Coil, A. L., Myers, A. D., Cox, T. J., & Spergel, D. N. 2007, ApJ, 662, 110
Hopkins, P. F., & Hernquist, L. 2009, ApJ, 694, 599
Ilbert, O., et al. 2005, A&A, 439, 863
Ilbert, O., et al. 2006, arXiv:astro-ph/0602329
Ilbert, O., Salvato, M., Le Floc'h, E., et al. 2010, ApJ, 709, 644
Ivezić, Ž., et al. 2004, Astronomische Nachrichten, 325, 583
Jester, S., Schneider, D. P., Richards, G. T., et al. 2005, AJ, 130, 873
Jing, Y. P., Mo, H. J., & Börner, G. 1998, ApJ, 494, 1
Kinney, A. L., Calzetti, D., Bohlin, R. C., McQuade, K., Storchi-Bergmann, T., & Schmitt, H. R. 1996, ApJ, 467, 38
Koutoulidis, L., Plionis, M., Georgantopoulos, I., & Fanidakis, N. 2013, MNRAS, 428, 1382
Kormendy, J., & Richstone, D. 1995, ARA&A, 33, 581
Kravtsov, A. V., Berlind, A. A., Wechsler, R. H., Klypin, A. A., Gottloeber, S., Allgood, B., & Primack, J. R. 2004, ApJ, 609, 35
Krumpe, M., Miyaji, T., & Coil, A. L. 2010, ApJ, 713, 558
Landy, S. D., & Szalay, A. S. 1993, ApJ, 412, 64
Laurikainen, E., & Salo, H. 1995, A&A, 293, 683
Lawrence, A., et al. 2007, MNRAS, 379, 1599
Le Fèvre, O., et al. 2004, A&A, 417, 839
Li, C., Kauffmann, G., Jing, Y. P., White, S. D. M., Börner, G., & Ch eng, F. Z. 2006, MNRAS, 368, 21
Li, C., Kauffmann, G., Wang, L., White, S. D. M., Heckman, T. M., & Jing, Y. P. 2007, The Central Engine of Active Galactic Nuclei, 373, 537
Loh, Y.-S., et al. 2010, MNRAS, 921
Lonsdale, C. J., Smith, H. E., Rowan-Robinson, M., et al. 2003, PASP, 115, 897 bibitem[Lu et al.(2006)]2006AJ....131..790L Lu, H., Zhou, H., Wang, J., et al. 2006, AJ, 131, 790
Lupton, R. H., Gunn, J. E., & Szalay, A. S. 1999, AJ, 118, 1406
Lupton, R., Gunn, J. E., Ivezić, Z., Knapp, G. R., & Kent, S. 2001, Astronomical Data Analysis Software and Systems X, 238, 269
Lutz, D., et al. 2010, ApJ, 712, 1287
Ma, B., Meng, K.-L., Pan, J., Huang, J.-S., & Feng, L.-L. 2009, Research in Astronomy and Astrophysics, 9, 979
Magliocchetti, M., et al. 2008, MNRAS, 383, 1131
Magorrian, J., et al. 1998, AJ, 115, 2285
Meneux, B., et al. 2008, A&A, 478, 299
Meneux, B., et al. 2009, A&A, 505, 463
Merritt, D., & Ferrarese, L. 2001, ApJ, 547, 140
Miyaji, T., et al. 2007, ApJS, 172, 396
Miyaji, T., Krumpe, M., Coil, A. L., & Aceves, H. 2010, arXiv:1010.5498
Mo, H. J., & White, S. D. M. 1996, MNRAS, 282, 347
Morrissey, P., et al. 2007, ApJS, 173, 682
Moster, B. P., Somerville, R. S., Newman, J. A., & Rix, H.-W. 2011, ApJ, 731, 113
Mountrichas, G., Sawangwit, U., Shanks, T., Croom, S. M., Schneider, D. P., Myers, A. D., & Pimbblet, K. 2009, MNRAS, 394, 2050
Mullaney, J. R., Alexander, D. M., Huynh, M., Goulding, A. D., & Frayer, D. 2010, MNRAS, 401, 995
Myers, A. D., et al. 2006, ApJ, 638, 622
Norberg, P., Baugh, C. M., Gaztañaga, E., & Croton, D. J. 2009, MNRAS, 396, 19
Norman, D. J., De Propris, R., & Ross, N. P. 2009, ApJ, 695, 1327
Padmanabhan, N., et al. 2008, ApJ, 674, 1217
Padmanabhan, N., White, M., Norberg, P., & Porciani, C. 2009, MNRAS, 397, 1862
Peacock, J. A., & Smith, R. E. 2000, MNRAS, 318, 1144
Peebles, P. J. E. 1980, The Large-Scale Structure of the Universe (Princeton, NJ: Princeton Univ. Press)
Pier, J. R., Munn, J. A., Hindsley, R. B., Hennessy, G. S., Kent, S. M., Lupton, R. H., & Ivezić, Ž. 2003, AJ, 125, 1559
Pollo, A., et al. 2006, A&A, 451, 409
Richards, G. T., Fan,X., Newberg, H. J., et al. 2002, AJ, 123, 2945
Schmidt, M. 1968, ApJ, 151, 393
Schneider, D. P., et al. 2007, AJ, 134, 102
Scoccimarro, R., Sheth, R. K., Hui, L., & Jain, B. 2001, ApJ, 546, 20
Scoville, N., et al. 2007, ApJS, 172, 1
Serber, W., Bahcall, N., Ménard, B., & Richards, G. 2006, ApJ, 643, 68
Shankar, F., Weinberg, D. H., & Shen, Y. 2010, MNRAS, 406, 1959
Shen, Y., et al. 2009, ApJ, 697, 1656
Shirasaki, Y., Tanaka, M., Ohishi, M., et al. 2011, PASJ, 63, 469
Small, T. A., & Blandford, R. D. 1992, MNRAS, 259, 725
Smith, R. J., Boyle, B. J., & Maddox, S. J. 1995, MNRAS, 277, 270
Smith, R. J., Boyle, B. J., & Maddox, S. J. 2000, MNRAS, 313, 252
Smith, J. A., et al. 2002, AJ, 123, 2121
Soltan, A. 1982, MNRAS, 200, 115
Sorrentino, G., Radovich, M., & Rifatto, A. 2006, A&A, 451, 809
Strand, N. E., Brunner, R. J., & Myers, A. D. 2008, ApJ, 688, 180
Stoughton, C., et al. 2002, AJ, 123, 485

Tucker, D. L., et al. 2006, Astronomische Nachrichten, 327, 821
Tinker, J., Kravtsov, A. V., Klypin, A., Abazajian, K., Warren, M., Yepes, G., Gottlöber, S., & Holz, D. E. 2008, ApJ, 688, 709
Ueda, Y., Akiyama, M., Ohta, K., & Miyaji, T. 2003, ApJ, 598, 886
Wang, J.-G., et al. 2009, ApJ, 707, 1334
Yang, Y., Mushotzky, R. F., Barger, A. J., & Cowie, L. L. 2006, ApJ, 645, 68
Yang, X. H., Mo, H. J., & van den Bosch, F. C. 2003, MNRAS, 339, 1057
Yee, H. K. C., & Green, R. F. 1984, ApJ, 280, 79
Yee, H. K. C., & Green, R. F. 1987, ApJ, 319, 28
York, D. G., et al. 2000, AJ, 120, 1579
Yu, Q., & Tremaine, S. 2002, MNRAS, 335, 965
Zehavi, I., et al. 2004, ApJ, 608, 16
Zehavi, I., et al. 2005, ApJ, 630, 1
Zheng, Z., et al. 2005, ApJ, 633, 791
Zheng, Z., Coil, A. L., & Zehavi, I. 2007, ApJ, 667, 760
Zheng, Z., Zehavi, I., Eisenstein, D. J., Weinberg, D. H., & Jing, Y. P. 2009, ApJ, 707, 554
Zucca, E., et al. 2006, A&A, 455, 879